\documentclass{aa}
\usepackage{graphicx}
\usepackage{bm}
\usepackage{times}
\graphicspath{{./fig/}{./png/}}

\newcommand{\EQ}{\begin{equation}}
\newcommand{\EE}{\end{equation}}
\newcommand{\EQA}{\begin{eqnarray}}
\newcommand{\EEA}{\end{eqnarray}}

\newcommand{\pd}{\partial}

\newcommand{\DIV}{\vec{\nabla} \cdot }
\newcommand{\CURL}{\vec{\nabla} \times }

\newcommand{\mean}[1]{\overline{#1}}
\newcommand{\meanv}[1]{\overline{\bm #1}}

\newcommand{\etat}{\eta_{\rm t}}
\newcommand{\etatz}{\eta_{\rm t0}}

\newcommand{\urms}{u_{\rm rms}}

\newcommand{\kef}{k_{\rm f}}
\newcommand{\tauc}{\tau_{\rm c}}

\newcommand{\St}{{\rm St}}
\newcommand{\Sh}{{\rm Sh}}
\newcommand{\Pm}{{\rm Pm}}
\newcommand{\Rm}{{\rm Rm}}
\newcommand{\Pra}{{\rm Pr}}
\newcommand{\Ra}{{\rm Ra}}

\newcommand{\Rey}{{\rm Re}}
\newcommand{\Co}{{\rm Co}}

\newcommand{\emf}{\bm{\mathcal{E}}}

{}
\def\onethird{{\textstyle{1\over3}}}
\def\onehalf{{\textstyle{1\over2}}}

\begin{document}

\authorrunning{K\"apyl\"a et al.}
\titlerunning{Alpha effect and turbulent diffusion from convection}

   \title{Alpha effect and turbulent diffusion from convection}

   \author{P. J. K\"apyl\"a
          \inst{1}
          \and
          M. J. Korpi
          \inst{1}
          \and
          A. Brandenburg
          \inst{2}
          }

   \offprints{\email{petri.kapyla@helsinki.fi}
              }

   \institute{Observatory, T\"ahtitorninm\"aki (PO Box 14), FI-00014 
              University of Helsinki, Finland
         \and NORDITA, AlbaNova University Center, Roslagstullsbacken
              23, SE-10691 Stockholm, Sweden}

   \date{Received 10 December 2008 / Accepted 26 March 2009}

   \abstract{}%
   {We study turbulent transport coefficients that describe the evolution of
     large-scale magnetic fields in turbulent convection.}%
   {We use the test field method, together with
     three-dimensional numerical simulations of turbulent convection
     with shear and rotation, to compute turbulent transport coefficients
     describing the evolution of large-scale magnetic fields in
     mean-field theory in the kinematic
     regime. We employ one-dimensional mean-field models with
     the derived turbulent transport coefficients to examine whether they
     give results that are compatible with direct simulations.}%
   {The results for the $\alpha$-effect as a function of rotation rate are
     consistent with earlier numerical studies, i.e.\ increasing
     magnitude as rotation increases and approximately $\cos \theta$
     latitude profile for moderate rotation.
     Turbulent diffusivity, $\etat$, is proportional to the square of the 
     turbulent vertical velocity in all cases. Whereas $\etat$ decreases 
     approximately inversely proportional to the wavenumber of the field, 
     the $\alpha$-effect and turbulent pumping show a more complex behaviour 
     with partial or full sign changes and the magnitude staying roughly 
     constant.
     In the presence of shear and no rotation, a weak $\alpha$-effect is induced which
     does not seem to show any consistent trend as a function of shear
     rate. Provided that the shear is large enough, this small
     $\alpha$-effect is able to excite a dynamo in the mean-field
     model. The coefficient responsible for driving the shear-current
     effect shows several sign changes as a function of depth but is
     also able to contribute to dynamo action in the mean-field model. The
     growth rates in these cases are, however, well below those in
     direct simulations, suggesting that an incoherent $\alpha$-shear
     dynamo may also act in the simulations.
     If both rotation and shear are present, the
     $\alpha$-effect is more pronounced. 
     At the same time, the combination
     of the shear-current and $\bm{\Omega}\times{\bm J}$-effects is
     also stronger than in the case of shear alone, but subdominant to 
     the $\alpha$-shear dynamo. The
     results of direct simulations are consistent with mean-field models
     where all of these effects are taken into account without the
     need to invoke incoherent effects.}%
   {}%


   \maketitle

\section{Introduction}
The solar magnetic field is thought to arise from a complicated
interplay of turbulence, rotation, and large-scale shear flows (e.g.\
Ossendrijver \cite{O03} and references therein).
Whilst numerical simulations of simple systems using fully
periodic boxes and externally forced idealised flows exhibiting
large-scale dynamos have been around for some time 
(e.g.\ Brandenburg
\cite{B01,B05a}; Brandenburg et al.\ \cite{BBS01};
Mininni et al.\ \cite{MGM05};
Brandenburg \& K\"apyl\"a \cite{BK07}; Yousef et al.\ \cite{YHSea08a,YHSea08b};
K\"apyl\"a \& Brandenburg
\cite{KB08}) and dynamos driven by the magnetorotational instability
exhibit large-scale dynamos (e.g.\ Brandenburg et al.\ \cite{BNST95}; Hawley et al.\ \cite{HGB96}),
convection simulations have not been able to produce appreciable
large-scale magnetic fields until recently (Rotvig \& Jones \cite{RJ02};
Browning et al.\ \cite{BMBT06};
Brown et al.\ \cite{BBBMNT07}; K\"apyl\"a et al.\
\cite{KKB08}, hereafter Paper I; Hughes \& Proctor \cite{HP08}). The
main ingredient missing in many earlier simulations was a large-scale
shear flow and boundary conditions which allow magnetic helicity
fluxes out of the system. Indeed, the shear flow plays a dual role in
dynamos: it not only generates new magnetic fields by stretching, but
it also drives magnetic helicity fluxes along constant isocontours of
shear which can allow efficient dynamo action (Vishniac \& Cho
\cite{VC01}; Brandenburg \& Subramanian \cite{BS05}; Paper I).
Recently, however, large-scale dynamos have also been found from rigidly
rotating convection simulations without shear 
(K\"apyl\"a et al.\ \cite{KKB08b}).

Although large-scale magnetic fields can clearly be obtained from
simulations, the origin of these fields in many cases (e.g.\ 
Yousef et al.\ \cite{YHSea08a,YHSea08b}; Paper I; 
Hughes \& Proctor \cite{HP08}) is still uncertain. In the
mean-field framework (e.g.\ Moffatt \cite{M78}; Parker
\cite{P79}; Krause \& R\"adler \cite{KR80}; R\"udiger \& Hollerbach \cite{RH04}), the
dynamo process is described by turbulent transport coefficients that
govern the evolution of large-scale magnetic field. 
The evolution equation for the large-scale part is obtained from the 
standard induction equation by decomposing magnetic and velocity 
fields into their mean and fluctuating parts, i.e.\ 
${\bm B}= \mean{\bm B} + {\bm b}$, ${\bm U}= \mean{\bm U} + {\bm u}$, which
leads to
\begin{equation}
\frac{\pd \mean{\bm B}}{\pd t} = \CURL({\mean{\bm U} \times \mean{\bm B} + \bm{\mathcal{\mean{E}}}} - \eta \mu_0 \mean{\bm J})\;, \label{equ:meanind}
\end{equation}
where $\eta$ is the molecular magnetic diffusivity, $\mean{\bm J} =
\mu_0^{-1}\CURL{\mean{\bm B}}$ is the current density, and $\mu_0$ is the
vacuum permeability.
The remaining term, $\mean\emf \equiv \overline{\bm{u} \times \bm{b}}$, is
the electromotive force describing the effects of small-scale
turbulence on the evolution of mean fields and can be represented in terms
of the mean fields and their derivatives
\begin{equation}
\mathcal{\mean{E}}_i = \alpha_{ij} \mean{B}_j + \eta_{ijk} \mean{B}_{j,k} + \ldots, \label{equ:emf}
\end{equation}
where $\alpha_{ij}$ and $\eta_{ijk}$ are tensorial coefficients, commas
denote partial derivatives, and summation over repeated indices is assumed.
Expression (\ref{equ:emf}) is valid if the mean fields vary slowly in space and
time.

Whilst mean-field models have been quite successful in
reproducing many aspects of the solar magnetism (e.g.\ Ossendrijver \cite{O03}), they have often been
hampered by the poor knowledge of the turbulent transport coefficients which could
only be computed analytically using unrealistic or unjustified
approximations, such as first order smoothing (FOSA).
More recently, numerical models of convection in local Cartesian
geometry have been employed to compute some of these coefficients in
more realistic setups (Brandenburg et al.\ \cite{BTNPS90};
Ossendrijver et al.\ \cite{OSB01}, \cite{OSRB02};
Giesecke et al.\ \cite{GZR05}; K\"apyl\"a et al.\ \cite{KKOS06};
Cattaneo \& Hughes \cite{CH06}; Hughes \& Cattaneo \cite{HC08}).
To date, however, only coefficients relevant for the
$\alpha_{ij}$ term in Eq.~(\ref{equ:emf})
have been determined from convection
simulations. This is due to the limitations of the method used where a
uniform magnetic field is imposed and the resulting electromotive force is
measured.
Furthermore, if the Lorentz force is retained in the simulations,
dynamo-generated magnetic fields may grow to saturation,
leading to quenching even if the imposed field is weak.
At large magnetic Reynolds numbers
such quenching can be very strong if there are no magnetic helicity fluxes,
suggesting therefore small values of $\alpha$ even for weak imposed fields.

During recent years an improved scheme of extracting turbulent transport
coefficients has appeared which is referred to as the test field method
(Schrinner et al.\ \cite{SRSRC05,Shea07}). In the test field method the
velocity field of the simulation is used in a number of induction
equations, which all correspond to a given set of large-scale test fields
which do neither evolve nor react back onto the velocity field.
The test fields are orthogonal so the coefficients can be obtained by
matrix inversion.
This method has been used successfully in setups where the
turbulence is due to isotropic forcing without shear (Sur et al.\
\cite{SBS08}, Brandenburg et al.\ \cite{BRS08}) and with shear
(Brandenburg et al.\ \cite{BRRK08}; Mitra et al.\ \cite{MKTB08}),
respectively. Moreover, the method has been used to extract dynamo
coefficients from more realistic setups where the turbulence is driven
by supernovae (Gressel et al.\ \cite{GZER08}) and the
magnetorotational instability (Brandenburg \cite{B05b,B08}). 

In the present paper we apply the method for the first time to
convection simulations. We also seek to understand the dynamos
reported in Paper I by applying the derived coefficients in a one-dimensional mean-field model. 
In the case of convection with rigid rotation it is likely that
the large-scale fields are due to the turbulent $\alpha$-effect that
is present in helical flows (K\"apyl\"a et al.\
\cite{KKB08b}). However, when shear is present, there are various
mechanisms that can generate large-scale fields: in helical flows a
finite $\alpha$-effect (e.g.\ R\"adler et al.\ \cite{RKR03}; R\"adler \& Stepanov \cite{RS06}; 
R\"udiger \& Kitchatinov \cite{RK06})
with shear can excite a classical
$\alpha\Omega$ or $\alpha$-shear-dynamo
(e.g.\ Brandenburg \& K\"apyl\"a \cite{BK07}; 
K\"apyl\"a \& Brandenburg \cite{KB08}).
Even if the mean value of $\alpha$ is zero,
strong enough fluctuations about zero in combination with shear can
drive an incoherent $\alpha$-shear dynamo (e.g.\ Vishniac \&
Brandenburg \cite{VB97}; Proctor \cite{P07}). Finally, the
shear-current (Rogachevskii \& Kleeorin \cite{RK03,RK04}; Kleeorin \& Rogachevskii \cite{KR08}) and
$\bm{\Omega}\times{\bm J}$ (R\"adler \cite{R69}; R\"adler et
al. \cite{RKR03}; Pipin \cite{P08}) effects may operate even in
nonhelical turbulence.
If both rotation and shear are present in the system it is
not obvious how to distinguish between the shear-current and 
$\bm{\Omega}\times{\bm J}$ effects.
In the present paper we
are able to extract the relevant turbulent transport coefficients responsible for
most of these processes and determine which one of them is dominant in
the different cases 
with the help of a one-dimensional mean-field model. In order to
facilitate comparisons between the mean-field models and the direct
simulations presented in Paper I, we use identical setups and
overlapping parameter regimes as those used in Paper I in the
determination of the transport coefficients.

\section{Model and methods}
The setup is similar to that used by, e.g., Brandenburg et al.\
(\cite{BJNRST96}), Ossendrijver et al.\ (\cite{OSB01}, \cite{OSRB02}),
and K\"apyl\"a et al.\ (\cite{KKT04}, \cite{KKOS06}) and in Paper I. A
small rectangular portion of a star is modelled by a box situated at
colatitude $\theta$. 
The coordinate system is such that $(x,y,z)$ corresponds to $(\theta,\phi,r)$
in a spherical coordinate system.
The dimensions of the domain are in most cases $(L_x, L_y, L_z)
= (4,4,2)d$, where $d$ is the depth of the convectively unstable
layer, and it is also used as our unit length. The box is divided
into three layers, an upper cooling layer, a convectively unstable
layer, and a stable overshoot layer (see below). The following set of
equations for compressible hydrodynamics is being solved:
\begin{equation}
\frac{\mathcal{D} \ln \rho}{\mathcal{D}t} = -\DIV{\bm U},
\end{equation}
\begin{equation}
\frac{\mathcal{D} \bm U}{\mathcal{D}t}\! = \! -S U_x \hat{\bm{y}} -\frac{1}{\rho}{\bm \nabla}p + {\bm g} -2\bm{\Omega}\times{\bm U} + \frac{1}{\rho} \bm{\nabla} \cdot 2 \nu \rho \bm{\mathsf{S}}, \label{equ:UU}
\end{equation}
\begin{equation}
\frac{\mathcal{D} \bm e}{\mathcal{D}t} = -\frac{p}{\rho}\DIV {\bm U} + \frac{1}{\rho} \bm{\nabla} \cdot K \bm{\nabla}T + 2 \nu \bm{\mathsf{S}}^2 - \frac{e-e_0}{\tau(z)}, \label{equ:ene}
\end{equation}
where $\mathcal{D}/\mathcal{D}t = \pd/\pd t + (\bm{U} +
\meanv{U}^{(S)}) \cdot \bm{\nabla}$, and $\meanv{U}^{(S)} =
(0,Sx,0)$ is the imposed large-scale shear flow. The kinematic
viscosity is given by $\nu$, $\rho$ is the density, $\bm{U}$ is the
velocity, and $\bm{g} = -g\hat{\bm{z}}$ is the gravitational acceleration.
The fluid obeys an ideal gas law $p=\rho e (\gamma-1)$, where $p$
and $e$ are the pressure and internal energy, respectively, and
$\gamma=c_{\rm P}/c_{\rm V} = 5/3$ is the ratio of specific heats in constant
pressure and volume. The internal energy is related to the temperature
via $e=c_{\rm V} T$, and $K$ is the heat conductivity.
The rate of strain tensor $\bm{\mathsf{S}}$ is given by
\begin{eqnarray}
\mathsf{S}_{ij} = \onehalf (U_{j,i}+U_{i,j}) - \onethird \delta_{ij} \DIV \bm{U}\;.
\end{eqnarray}
The last term of Eq.~(\ref{equ:ene}) describes cooling at the top of
the domain, where $\tau(z)$ is a cooling time which has a profile
smoothly connecting the upper cooling layer and the convectively
unstable layer below.

The coordinates $(z_1, z_2, z_3, z_4) = (-0.85, 0, 1, 1.15)d$ give the
vertical positions of the bottom of the box, the bottom and top of the
convectively unstable layer, and the top of the box,
respectively.
We use a $K(z)$ profile such that the associated hydrostatic reference
solution is piecewise polytropic with indices $(m_1, m_2, m_3)=(3, 1, 1)$.
The cooling layer near the top makes that layer nearly isothermal
and hence stably stratified.
The bottom layer is also stably stratified, and the middle layer is
convectively unstable.

Stress-free boundary conditions are used for the velocity,
\begin{equation}
U_{x,z} = U_{y,z} = U_z = 0.
\end{equation}
In the absence of shear the $x$ and $y$ directions are periodic whereas
if shear is present, shearing-periodic conditions are used in the $x$ 
direction.
A constant temperature gradient is maintained at the bottom of the box which leads to
a steady influx of heat due to the constant heat conductivity. 
The simulations were made with the {\sc Pencil Code}%
\footnote{\texttt{http://www.nordita.org/software/pencil-code/}},
which uses sixth-order explicit finite differences in space and third
order accurate time stepping method.
Resolutions of up to $256^3$ mesh points were used.

\subsection{Units, nondimensional quantities, and parameters}

Dimensionless quantities are obtained by setting
\begin{eqnarray}
d = g = \rho_0 = c_{\rm P} = \mu_0 = 1\;,
\end{eqnarray}
where $\rho_0$ is the density at $z_2$. The units of length, time,
velocity, density, entropy, and magnetic field are then
\begin{eqnarray}
&& [x] = d\;,\;\; [t] = \sqrt{d/g}\;,\;\; [U]=\sqrt{dg}\;,\;\; [\rho]=\rho_0\;,\;\; \nonumber \\ && [s]=c_{\rm P}\;,\;\; [B]=\sqrt{dg\rho_0\mu_0}\;. 
\end{eqnarray}
The simulations are then governed by the dimensionless numbers
\begin{eqnarray}
\Pra=\frac{\nu}{\chi_0}\;,\;\; \Rey=\frac{\urms}{\nu \kef}  \;,\;\; \Ra=\frac{gd^4}{\nu \chi_0} \bigg(-\frac{1}{c_{\rm P}}\frac{{\rm d}s}{{\rm d}z} \bigg)_{z_{\rm m}}\;,
\end{eqnarray}
where $\chi_0 = K/(\rho_{\rm m} c_{\rm P})$ is the thermal
diffusivity, $\kef=2\pi/d$ is an estimate of the wavenumber of the
energy-carrying eddies, and $\rho_{\rm m}$ is the density in the middle of
the unstable layer at \ $z_{\rm m}=\onehalf(z_3-z_2)$.
Our choice of $\kef$ is somewhat arbitrary because it is difficult to
define a single length scale which would describe the flow in a
highly inhomogeneous system such as stratified convection. The
vertical extent of convective cells, however, is almost always of the
order of the depth of the convectively unstable layer which suggests
that $d$ could be used as the length scale describing convection. In
the nonrotating case, this is also close to the horizontal size of the
convective eddies.
The entropy gradient, measured at $z_{\rm m}$ in the non-convecting
initial state, is given by
\begin{eqnarray}
\bigg(-\frac{1}{c_{\rm P}}\frac{{\rm d}s}{{\rm d}z}\bigg)_{z_{\rm m}} = \frac{\nabla-\nabla_{\rm ad}}{H_{\rm P}}\;,
\end{eqnarray}
with $\nabla_{\rm ad} = 1-1/\gamma$ and $\nabla = (\pd \ln
T/\pd \ln p)_{z_{\rm m}}$, and $H_{\rm P}$ is the pressure scale height.

The amount of stratification is determined by the parameter
\begin{eqnarray}
\xi_0 = \frac{(\gamma-1) e_0}{gd}\;,
\end{eqnarray}
where $e_0$ is the internal energy at $z_4$. 
We use $\xi_0=1/3$ in all models.

\begin{table*}[t]
\centering
\caption[]{Summary of the runs. The numbers are given for the statistically 
  saturated state. Here, $\tilde{k}=k/k_1$, $\mbox{Ma}=\urms/(gd)^{1/2}$, and
  $L_{\rm H}=L_x=L_y$.}
      \label{tab:runs}
      \vspace{-0.5cm}
     $$
         \begin{array}{p{0.035\linewidth}cccccccccccc}
           \hline
           \noalign{\smallskip}
Run & $grid$ & $\Pra$ & $\Ra$ & $Rm$ & \Pm & \Sh & \Co & \theta & \mbox{Ma} & \tilde{k} & L_{\rm H} \\ \hline 
A   & 128^3  & 1.37 & 3.1\cdot10^5 & 37  & 5 &     0  & 0    &    -    & 0.046 & 1 & 4 \\ 
B   & 128^3  & 1.37 & 3.1\cdot10^5 & 35  & 5 &     0  & 0.36 &  0\degr & 0.043 & 1 & 4 \\ 
C   & 128^3  & 1.37 & 3.1\cdot10^5 & 46  & 5 & -0.14  & 0    &    -    & 0.058 & 1 & 4 \\ 
D   & 128^3  & 1.37 & 3.1\cdot10^5 & 35  & 5 & -0.18  & 0.36 &  0\degr & 0.044 & 1 & 4 \\ 
\hline
A1   & 128^3  & 1.37 & 3.1\cdot10^5 & 37  & 5 &    0  & 0    &    -    & 0.046 & 0 & 4 \\ 
A2   & 128^3  & 1.37 & 3.1\cdot10^5 & 37  & 5 &    0  & 0    &    -    & 0.046 & 1 & 4 \\ 
A3   & 128^3  & 1.37 & 3.1\cdot10^5 & 38  & 5 &    0  & 0    &    -    & 0.048 & 2 & 4 \\ 
A4   & 128^3  & 1.37 & 3.1\cdot10^5 & 38  & 5 &    0  & 0    &    -    & 0.048 & 3 & 4 \\ 
           \hline
B1   & 64\times128^2  & 1.37 & 3.1\cdot10^5 & 33  & 5 &     0  & 0.38 &  0\degr & 0.042 & 1 & 2 \\ 
B2   & 128^3          & 1.37 & 3.1\cdot10^5 & 35  & 5 &     0  & 0.36 &  0\degr & 0.043 & 1 & 4 \\ 
B3   & 256\times128^2 & 1.37 & 3.1\cdot10^5 & 32  & 5 &     0  & 0.40 &  0\degr & 0.040 & 1 & 8 \\ 
           \hline
B4   & 128^3  & 0.69 & 6.1\cdot10^5 & 1.6 & 0.1&     0  & 0.32 &  0\degr & 0.049 & 1 & 4 \\ 
B5   & 128^3  & 0.69 & 6.1\cdot10^5 & 3.2 & 0.2&     0  & 0.32 &  0\degr & 0.050 & 1 & 4 \\ 
B6   & 128^3  & 0.69 & 6.1\cdot10^5 & 7.8 & 0.5&     0  & 0.33 &  0\degr & 0.049 & 1 & 4 \\ 
B7   & 128^3  & 0.69 & 6.1\cdot10^5 & 16  & 1  &     0  & 0.32 &  0\degr & 0.050 & 1 & 4 \\ 
B8   & 128^3  & 0.69 & 6.1\cdot10^5 & 32  & 2  &     0  & 0.32 &  0\degr & 0.050 & 1 & 4 \\ 
B9   & 128^3  & 0.69 & 6.1\cdot10^5 & 75  & 5  &     0  & 0.34 &  0\degr & 0.047 & 1 & 4 \\ 
B10  & 256^3  & 0.69 & 6.1\cdot10^5 & 155 & 10 &     0  & 0.33 &  0\degr & 0.049 & 1 & 4 \\ 
           \hline
B11  & 128^3  & 1.37 & 3.1\cdot10^5 & 35  & 5 &     0  & 0.36 &  0\degr & 0.044 & 0 & 4 \\ 
B12  & 128^3  & 1.37 & 3.1\cdot10^5 & 35  & 5 &     0  & 0.36 &  0\degr & 0.043 & 1 & 4 \\ 
B13  & 128^3  & 1.37 & 3.1\cdot10^5 & 35  & 5 &     0  & 0.36 &  0\degr & 0.045 & 2 & 4 \\ 
B14  & 128^3  & 1.37 & 3.1\cdot10^5 & 35  & 5 &     0  & 0.36 &  0\degr & 0.045 & 3 & 4 \\ 
           \hline
B15  & 128^3  & 1.37 & 3.1\cdot10^5 & 35  & 5 &     0  & 0.07 &  0\degr & 0.044 & 1 & 4 \\ 
B16  & 128^3  & 1.37 & 3.1\cdot10^5 & 33  & 5 &     0  & 0.15 &  0\degr & 0.042 & 1 & 4 \\ 
B17  & 128^3  & 1.37 & 3.1\cdot10^5 & 35  & 5 &     0  & 0.36 &  0\degr & 0.044 & 1 & 4 \\ 
B18  & 128^3  & 1.37 & 3.1\cdot10^5 & 33  & 5 &     0  & 0.78 &  0\degr & 0.041 & 1 & 4 \\ 
B19  & 128^3  & 1.37 & 3.1\cdot10^5 & 29  & 5 &     0  & 1.74 &  0\degr & 0.037 & 1 & 4 \\ 
B20  & 128^3  & 1.37 & 3.1\cdot10^5 & 20  & 5 &     0  & 6.43 &  0\degr & 0.025 & 1 & 4 \\ 
           \hline
B21  & 128^3  & 1.37 & 3.1\cdot10^5 & 35  & 5 &     0  & 0.36 &  0\degr & 0.044 & 1 & 4 \\ 
B22  & 128^3  & 1.37 & 3.1\cdot10^5 & 36  & 5 &     0  & 0.35 & 15\degr & 0.045 & 1 & 4 \\ 
B23  & 128^3  & 1.37 & 3.1\cdot10^5 & 37  & 5 &     0  & 0.34 & 30\degr & 0.046 & 1 & 4 \\ 
B24  & 128^3  & 1.37 & 3.1\cdot10^5 & 38  & 5 &     0  & 0.33 & 45\degr & 0.048 & 1 & 4 \\ 
B25  & 128^3  & 1.37 & 3.1\cdot10^5 & 37  & 5 &     0  & 0.34 & 60\degr & 0.047 & 1 & 4 \\ 
B26  & 128^3  & 1.37 & 3.1\cdot10^5 & 40  & 5 &     0  & 0.32 & 75\degr & 0.050 & 1 & 4 \\ 
B27  & 128^3  & 1.37 & 3.1\cdot10^5 & 43  & 5 &     0  & 0.29 & 90\degr & 0.054 & 1 & 4 \\ 
           \hline
C1   & 128^3  & 1.37 & 3.1\cdot10^5 & 43  & 5 & -0.03 & 0     &    -    & 0.054 & 1 & 4 \\ 
C2   & 128^3  & 1.37 & 3.1\cdot10^5 & 42  & 5 & -0.06 & 0     &    -    & 0.052 & 1 & 4 \\ 
C3   & 128^3  & 1.37 & 3.1\cdot10^5 & 46  & 5 & -0.14 & 0     &    -    & 0.058 & 1 & 4 \\ 
C4   & 128^3  & 1.37 & 3.1\cdot10^5 & 66  & 5 & -0.19 & 0     &    -    & 0.083 & 1 & 4 \\ 
           \hline
D1   & 128^3  & 1.37 & 3.1\cdot10^5 & 37  & 5 & -0.03 & 0.06  &  0\degr & 0.046 & 1 & 4 \\ 
D2   & 128^3  & 1.37 & 3.1\cdot10^5 & 37  & 5 & -0.07 & 0.15  &  0\degr & 0.043 & 1 & 4 \\ 
D3   & 128^3  & 1.37 & 3.1\cdot10^5 & 37  & 5 & -0.18 & 0.36  &  0\degr & 0.044 & 1 & 4 \\ 
D4   & 128^3  & 1.37 & 3.1\cdot10^5 & 37  & 5 & -0.36 & 0.73  &  0\degr & 0.044 & 1 & 4 \\ 
D5   & 128^3  & 1.37 & 3.1\cdot10^5 & 37  & 5 & -0.83 & 1.66  &  0\degr & 0.038 & 1 & 4 \\ 
           \hline
         \end{array}
     $$ 
\end{table*}

\subsection{The test field method}
We employ the test field method (Schrinner et al.\ \cite{SRSRC05,Shea07}), 
which is implemented into the {\sc Pencil Code}, 
to determine turbulent transport coefficients.
The uncurled induction equation in
the shearing box approximation can be written
in terms of the vector potential in the Weyl gauge as
\begin{equation}
\frac{D \bm A}{Dt} = -S A_y \hat{\bm{x}} + \bm{U} \times \bm{B} - \eta \mu_0 {\bm J}, \label{equ:AA}
\end{equation}
where $D/Dt = \pd/\pd t + Sx\pd/\pd y$, $\bm{A}$ is the magnetic
vector potential, and $\bm{B} = \bm{\nabla} \times \bm{A}$ is the 
magnetic field.
The relative importance of magnetic diffusion over viscous and inertial
forces can be characterized respectively in terms of
magnetic Prandtl and Reynolds numbers
\begin{eqnarray}
\Pm=\frac{\nu}{\eta}, \;\;\; \Rm\equiv\frac{\urms}{\eta \kef} = \Pm\, \Rey. 
\end{eqnarray}
In most cases we use $\Pm=5$ and $\Rm\approx35$, see Table \ref{tab:runs}. When we vary $\Rm$ in 
the range from roughly 1.5 to 150, we keep $\Rey\approx15$ and vary $\Pm$ in 
the range $0.1-10$.
We decompose the fields
into their mean and fluctuating parts according to
\begin{equation}
\bm{A} = \meanv{A} + \bm{a}, \;\;
\bm{U} = \meanv{U} + \bm{u}, \;\;
\bm{B} = \meanv{B} + \bm{b}, \;\;
\bm{J} = \meanv{J} + \bm{j},
\end{equation}
where the overbars denote a horizontal average and lowercase
quantities denote fluctuations around these averages.
The equation for the mean vector potential is then
\begin{equation}
\frac{D \meanv{A}}{Dt} = -S \mean{A}_y \hat{\bm{x}} - \meanv{U} \times \meanv{B} + \mean{\bm{u} \times \bm{b}} - \eta \mu_0 \meanv{J}. \label{equ:AAmean}
\end{equation}
Subtracting (\ref{equ:AAmean}) from (\ref{equ:AA}) gives an equation
for the fluctuating field which reads
\begin{equation}
\frac{D \bm{a}}{Dt} = -S a_y \hat{\bm{x}} + \meanv{U} \times \bm{b} + \bm{u} \times \meanv{B} + \bm{u} \times \bm{b} - \mean{\bm{u} \times \bm{b}} - \eta \mu_0 \bm{j}. \label{equ:AAfluc}
\end{equation}
Instead of using the actual mean fields $\meanv{B}$ in this
equations, they are replaced by orthogonal test fields
$\meanv{B}^{p,q}$ and a separate Eq.~(\ref{equ:AAfluc}) is solved for
each one of them. Here we follow the same procedure as in
Brandenburg et al.\ (\cite{BRRK08}) and Mitra et al.\ (\cite{MKTB08})
and limit the study to mean magnetic fields that depend on $z$ only.
We use test fields
\begin{eqnarray}
\meanv{B}^{1c}&=&B_0(\cos kz,0,0), \quad \meanv{B}^{2c}=B_0(0,\cos kz,0),\\
\meanv{B}^{1s}&=&B_0(\sin kz,0,0), \quad \meanv{B}^{2s}=B_0(0,\sin kz,0),
\end{eqnarray}
where $k$ is the wavenumber of the test field. In most models 
we use $k/k_1=1$, where $k_1=2\pi/L_z$.
The electromotive force can be written as
\begin{equation}
\mean{\mathcal{E}}_i=\alpha_{ij}\mean{B}_j -\eta_{ij}\mu_0\mean{J}_j,
\end{equation}
where $\eta_{i1}= \eta_{i23}$ and $\eta_{i2}= -\eta_{i13}$. The 4+4
coefficients are then obtained by inverting a simple matrix equation,
relating the rank-2 tensor components to rank-3 tensor components.

Owing to the use of periodic boundary conditions in the horizontal
directions, the $z$-component of the mean magnetic field is conserved
and equal to the initial value, i.e.\ $\mean{B}_3=0$.
Therefore the value of $\alpha_{33}$ is here of no interest.

It is convenient to discuss the results in terms of the quantities
\begin{eqnarray}
\gamma&=&\onehalf(\alpha_{21}-\alpha_{12}), \ \ \epsilon_\gamma=\onehalf(\alpha_{21}+\alpha_{12}), \\ \etat&=&\onehalf(\eta_{11}+\eta_{22}), \ \ \epsilon_\eta=\onehalf(\eta_{11}-\eta_{22}), \\ \delta&=&\onehalf(\eta_{21}-\eta_{12}).
\end{eqnarray}
Furthermore, the remaining or otherwise important coefficients are
analyzed individually.
The most important of these are the diagonal components of
$\alpha_{ij}$ and $\eta_{21}$. The
former are responsible for the generation of magnetic fields in
helical turbulence and the latter
can drive the mean-field shear-current dynamo
in nonhelical turbulence with shear (Rogachevskii \& Kleeorin
\cite{RK03}, \cite{RK04}).

To normalize our results, we use isotropic expressions of $\alpha$ and
$\etat$ as obtained from first order smoothing, i.e.\
\begin{eqnarray}
\alpha_0 = \onethird \urms, \quad \etatz=\onethird \urms \kef^{-1},
\end{eqnarray}
where the root mean square velocity is a volume average and the Strouhal
number,
\begin{equation}
\St = \tauc \urms \kef,
\end{equation}
has been assumed to be of the order of unity.
In order to actually compare our results with those of FOSA,
anisotropic expressions need to be used. Such expressions have been
computed in the past (e.g.\ R\"adler \cite{R80}; see also K\"apyl\"a
et al.\ \cite{KKOS06}) and are given for the $\alpha$-effect, $\gamma$,
and $\etat$ by
\begin{eqnarray}
\alpha_{xx}^{(0)} &=& -2\tauc \mean{u_z \pd_x u_y},\label{equ:alpxx}\\
\alpha_{yy}^{(0)} &=& -2\tauc \mean{u_x \pd_y u_z},\label{equ:alpyy}\\
\gamma^{(0)} &=& -\tauc \pd_z \mean{u_z^2}, \label{equ:gamma0}\\
\etatz^{(0)} &=& \tauc \mean{u_z^2}, \label{equ:etat0}
\end{eqnarray}
where we have used integration by parts and assumed that $\tauc$ does not
depend on spatial coordinates.
The correlation time can be presented in terms $\urms$ and $\kef$ by
assuming a value for $\St$.

\subsection{Averaging and error estimates}
\label{sec:averr}
In the present study a mean quantity is considered to be a horizontal
average, defined via
\begin{equation}
\meanv{F}=\frac{1}{L_xL_y}
\int_{-{1\over2}L_y}^{{1\over2}L_y} \int_{-{1\over2}L_x}^{{1\over2}L_x}
    \bm{F}(x + x', y + y', z, t) \, \mbox{d} x' \mbox{d} y'.
\label{average}
\end{equation}
Except for special terms such as the shear terms in Eqs.~(\ref{equ:UU})
and (\ref{equ:AA}), this formulation corresponds to simple horizontal
averaging (for details see Brandenburg et al.\ \cite{BRRK08}).
An additional time average over the statistically steady part of each
simulations is also applied. The fluctuating magnetic fields
$\bm{b}^{p,q}$ are reset to zero after periodic time intervals in
order to avoid the complications arising from the growth of these fields;
see the more thorough discussions in Sur et al.\ (\cite{SBS08}) and
Mitra et al.\ (\cite{MKTB08}).

We estimate errors by computing the standard deviation $\sigma$ for
each depth and dividing this by the square root of the number of
independent realizations $N$ of the dynamo coefficients. We consider
the time series between two resets of the field $\bm{b}^{p,q}$ to
represent an independent realization. For a typical run, $N$ is between five
and ten.

\subsection{Corresponding mean-field models}
In order to determine how well the derived dynamo coefficients
describe the dynamos seen in direct simulations of Paper I, we
construct a one-dimensional mean-field model where the test field results
can be used directly as inputs. We start from the mean-field induction
equation, Eq.~(\ref{equ:meanind}), which can be written using the
vector potential
\begin{eqnarray}
\dot{\mean{A}}_i=-\mean{U}_{j,i}\mean{A}_j+\alpha_{ij}\mean{B}_j
-(\eta_{ij}+\eta\delta_{ij})\mu_0\mean{J}_j
\end{eqnarray}
where the dot on $\dot{\mean{A}}_i$ denotes a time derivative and
$\mean{U}_{j,i}\mean{A}_j=(S\mean{A}_y,0,0)$ is the shear term.
The mean magnetic field is given by
$\mean{\bm{B}}=(-\mean{A}_y',\mean{A}_x',0)$,
the mean current density is given by $\mu_0\mean{J}_i=-\mean{A}_i''$,
and primes denote $z$-derivatives.
The coefficients $\alpha_{ij}$ and $\eta_{ij}$ are taken directly from
the test field results leaving little freedom in the model.
We can, however, turn on and off any component of $\alpha_{ij}$ and
$\eta_{ij}$ when needed in order to study the effects of the different 
coefficients individually.

\begin{figure}[t]
\centering
\includegraphics[width=\columnwidth]{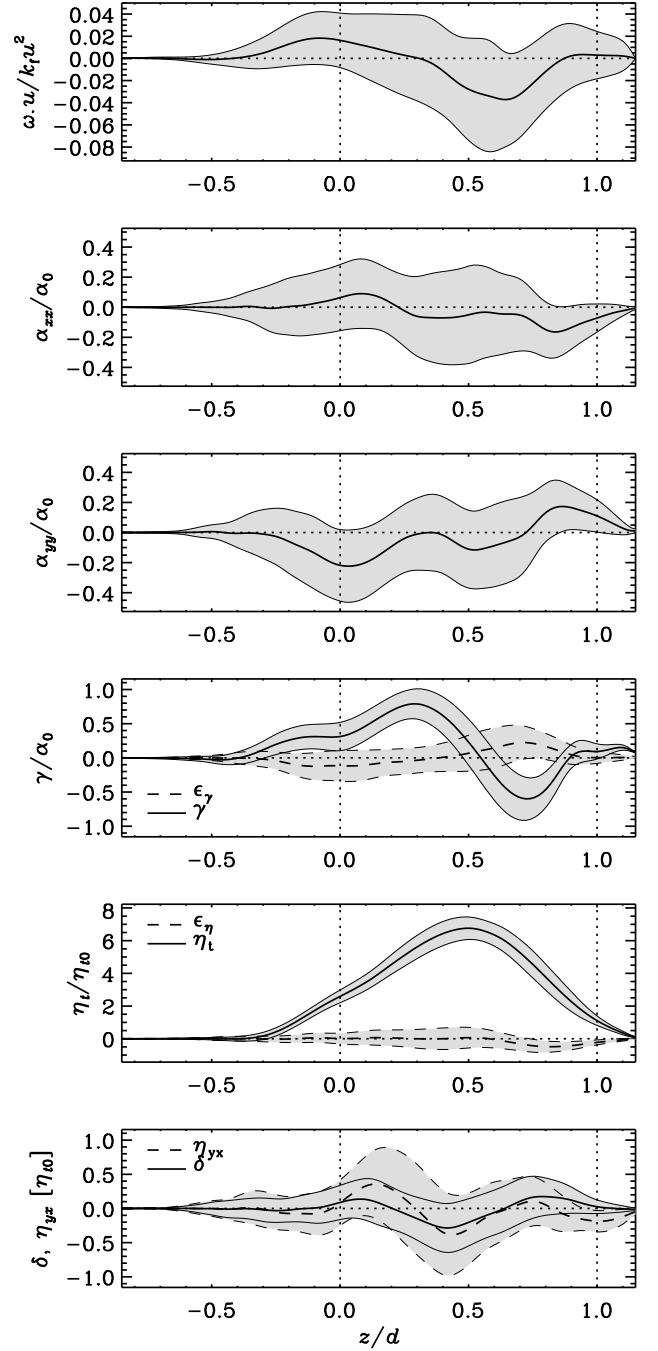}
\caption{
The three topmost panels show the time-averaged vertical profiles of 
kinetic helicity, $\alpha_{xx}$, and $\alpha_{yy}$, respectively.
The fourth and fifth panels show $\gamma$ with $\epsilon_\gamma$, and 
$\etat$ with $\epsilon_\eta$, respectively. The lowermost panel shows 
$\delta$ (solid line) and $\eta_{yx}$ (dashed).
From Run~A with $\Co=\Sh=0$, and $\Rm\approx37$.
The shaded areas between the thinner lines indicate error estimates 
as described in Sect.~\ref{sec:averr}.
The vertical lines at $z=(0,d)$ denote the base and top of the
convectively unstable layer.}
\label{fig:aenorot}
\end{figure}

\begin{figure}[t]
\centering
\includegraphics[width=\columnwidth]{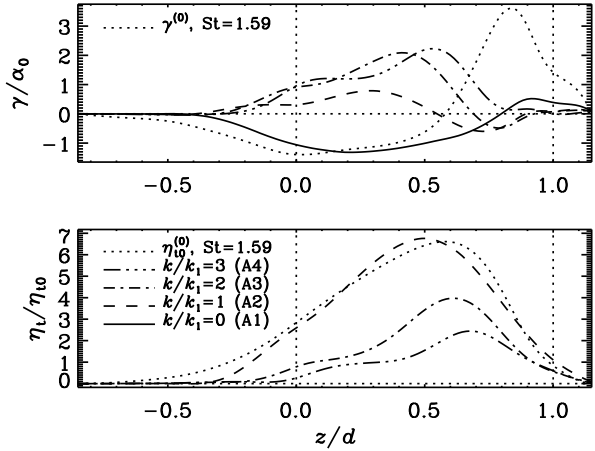}
\caption{Coefficients $\gamma$ (top panel) and $\etat$ (bottom panel)
  as functions of $k$ from Runs~A1--A4 with $\Rm\approx37$, and
  $\Co=\Sh=0$. The dotted lines show the FOSA results for $\gamma$ and
  $\eta$, according to Eqs.~(\ref{equ:gamma0}) and (\ref{equ:etat0}),
  respectively, with $\St\approx1.59$.}
\label{fig:kdep_128a}
\end{figure}

\section{Results}
In a similar fashion as in Paper I we perform four types of
simulations which we label as follows: in set~A neither rotation nor
shear is present whereas in set~B rotation is added. In set~C only
shear is present, and finally in set~D both rotation and shear are
used. Parameters such as the strengths of rotation and shear, as
measured by $\Co$ and $\Sh$, respectively, are varied within each set
to probe the parameter space. Summary of the runs is presented in Table \ref{tab:runs}.

The fluid Reynolds numbers in our simulations are quite modest so we
cannot consider our flows to be highly turbulent. However, the flows are
irregular enough to remain time dependent in all cases, as can also
be seen from various animations\footnote{%
\texttt{http://www.helsinki.fi/\ensuremath{\sim}kapyla/movies.html}}.

\subsection{Set A: no rotation nor shear  $(\Co=\Sh=0)$}
The simplest case we can consider with the present setup is one
with no rotation and no shear. 
In that case no net helicity
generation or $\alpha$-effect are expected. However, due to the
density stratification, the turbulence is inhomogeneous. This can lead
to a non-zero pumping, or $\gamma$-effect (e.g.\ Krause \& R\"adler
\cite{KR80}).

The horizontally and temporally averaged transport coefficients from
Run~A with $\Co=\Sh=0$ and $\Rm\approx37$ are presented in
Fig.~\ref{fig:aenorot}. The results show that the kinetic helicity 
is small and that the mean values of the diagonal elements of 
$\alpha_{ij}$ are of the order of $0.1\alpha_0$ with errors clearly larger than 
the mean. Vanishing diagonal elements of $\alpha_{ij}$ is
in accordance with expectations from symmetry arguments.
There is however a non-zero pumping effect directed upward (downward) in
the lower (upper) part of the convectively unstable layer. The sign of
the pumping is inconsistent with the diamagnetic effect, i.e.\ 
$\gamma \propto -\partial_z \mean{u_z^2}$ (e.g.\
R\"adler \cite{R68}) and differs from earlier results from convection
simulations using the imposed field method (Ossendrijver et al.\
\cite{OSRB02}; K\"apyl\"a et al.\ \cite{KKOS06})
and other diagnostics (e.g.\ Nordlund et al.\ \cite{NBJRRST92};
Tobias et al.\ \cite{TBCT98,TBCT01}; Ziegler \& R\"udiger \cite{ZR03}).
However, this result is obtained for test fields for which $k/k_1=1$,
whereas the imposed field results use a uniform field with
$k/k_1=0$. For a uniform test field the pumping effect indeed changes
sign and is thus consistent with the earlier numerical studies and the 
diamagnetic effect (see the upper panel
of Fig.~\ref{fig:kdep_128a}).
The FOSA-prediction, Eq.~(\ref{equ:gamma0}) for the turbulent pumping is in qualitative
agreement with the simulation result for $k/k_1=0$ but opposite to 
the results for $k/k_1$ greater than that.

At first glance the magnitude of the turbulent diffusivity seems quite high: the
maximum value is more than six times the isotropic reference value
$\etatz$ suggesting that $\St\approx6$.
However, the high value of $\etat$ turns out to be related to the
normalization: if an anisotropic expression, i.e.\
Eq.~(\ref{equ:etat0}), is plotted alongside $\etat$ the Strouhal
number is roughly 1.6, not six, for our standard case $k/k_1=1$, 
see the lower panel of
Fig.~(\ref{fig:kdep_128a}). The profile of the turbulent diffusivity
coincides with that of the vertical velocity squared as predicted by
Eq.~(\ref{equ:etat0}).
When $k$ is increased, the profile of $\etat$ stays roughly the same
and the magnitude diminishes roughly in proportion to $k^{-1}$.
 The quantities $\epsilon_\gamma$, $\epsilon_\eta$, $\delta$,
$\eta_{xy}$, and $\eta_{yx}$ are compatible with zero in all runs in set~A.

\begin{figure}[t]
\centering
\includegraphics[width=\columnwidth]{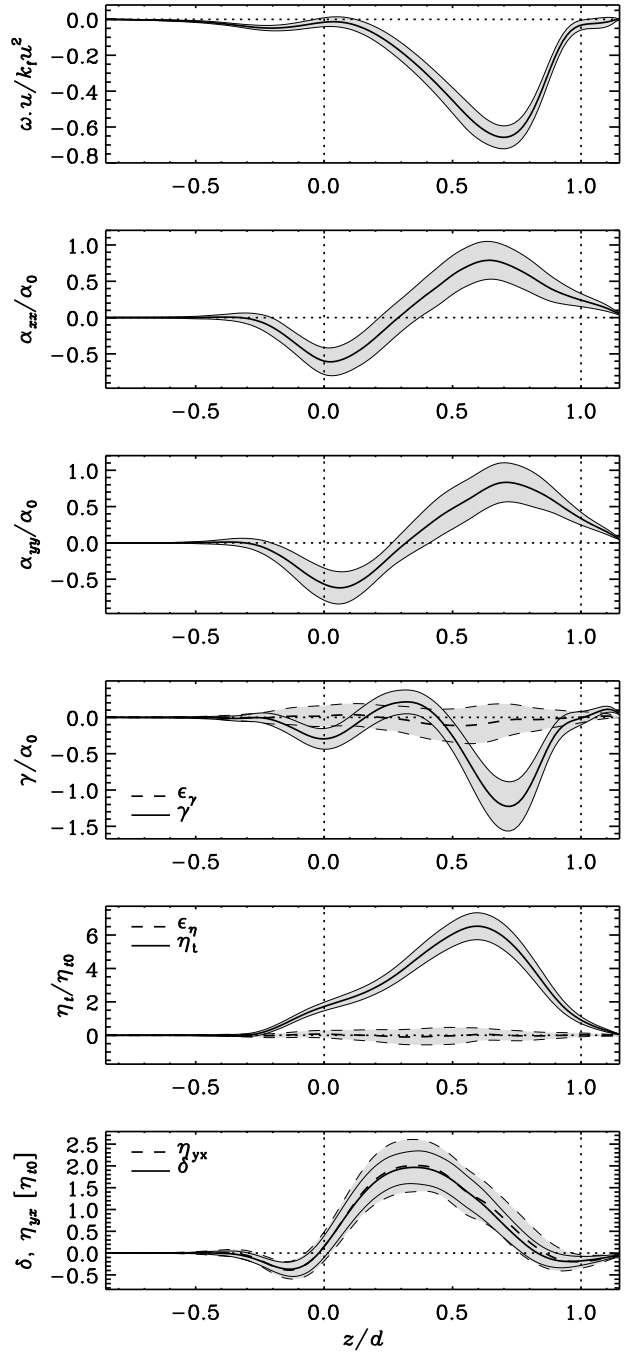}
\caption{Same as Fig.~\ref{fig:aenorot}, but for Run~B;
  $\Co\approx0.36$, $\theta=\Sh=0$, and $\Rm\approx35$.}
\label{fig:aerot}
\end{figure}

\begin{figure}[t]
\centering
\includegraphics[width=\columnwidth]{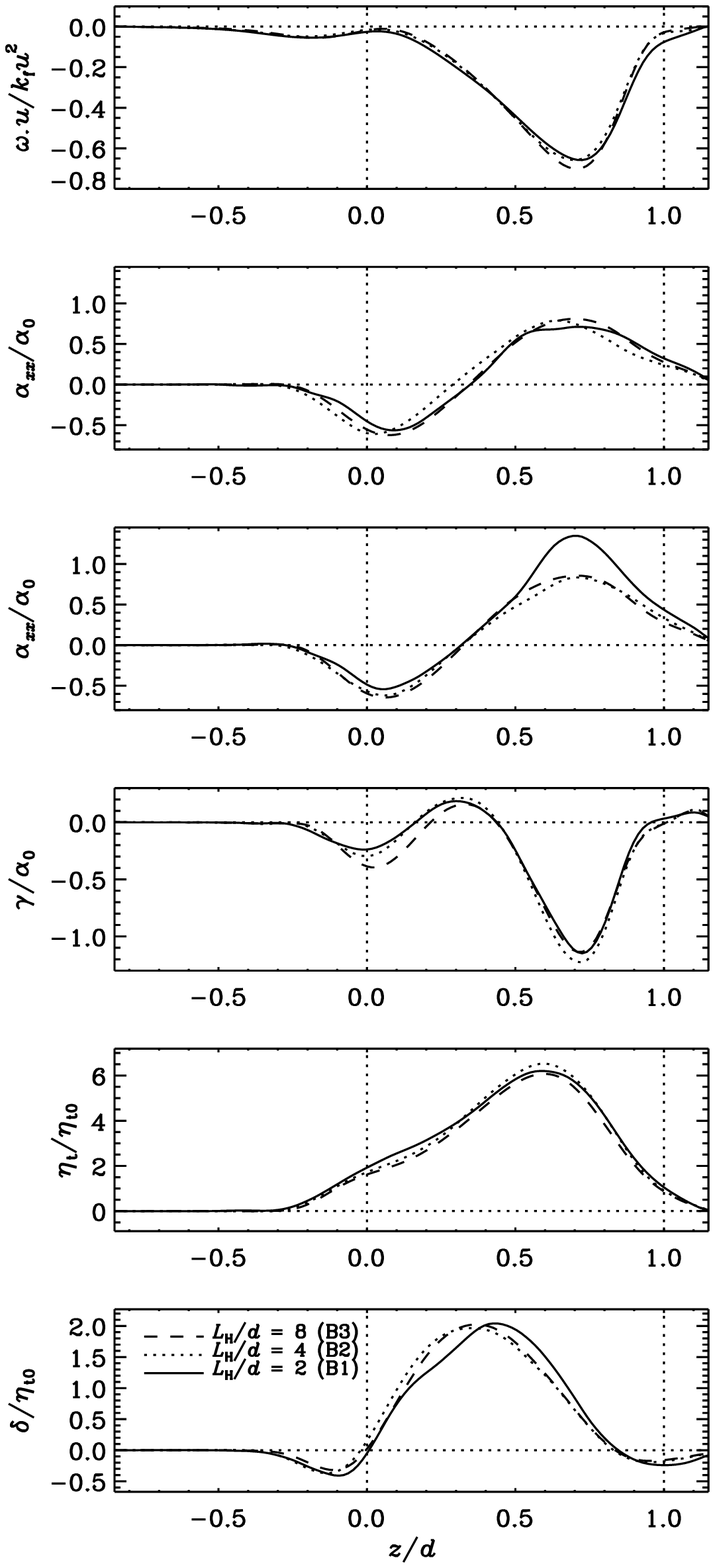}
\caption{From top to bottom: kinetic helicity, $\alpha_{xx}$,
  $\alpha_{yy}$, $\gamma$, $\etat$, and $\delta$ as functions of
  horizontal system size from Runs~B1--B3. The linestyles are as indicated in the lowermost
  panel. $\Co\approx0.36$, $\theta=0$, $\Sh=0$ and $\Rm\approx35$ in
  all runs.}
\label{fig:palpsize}
\end{figure}

\begin{figure}[t]
\centering
\includegraphics[width=\columnwidth]{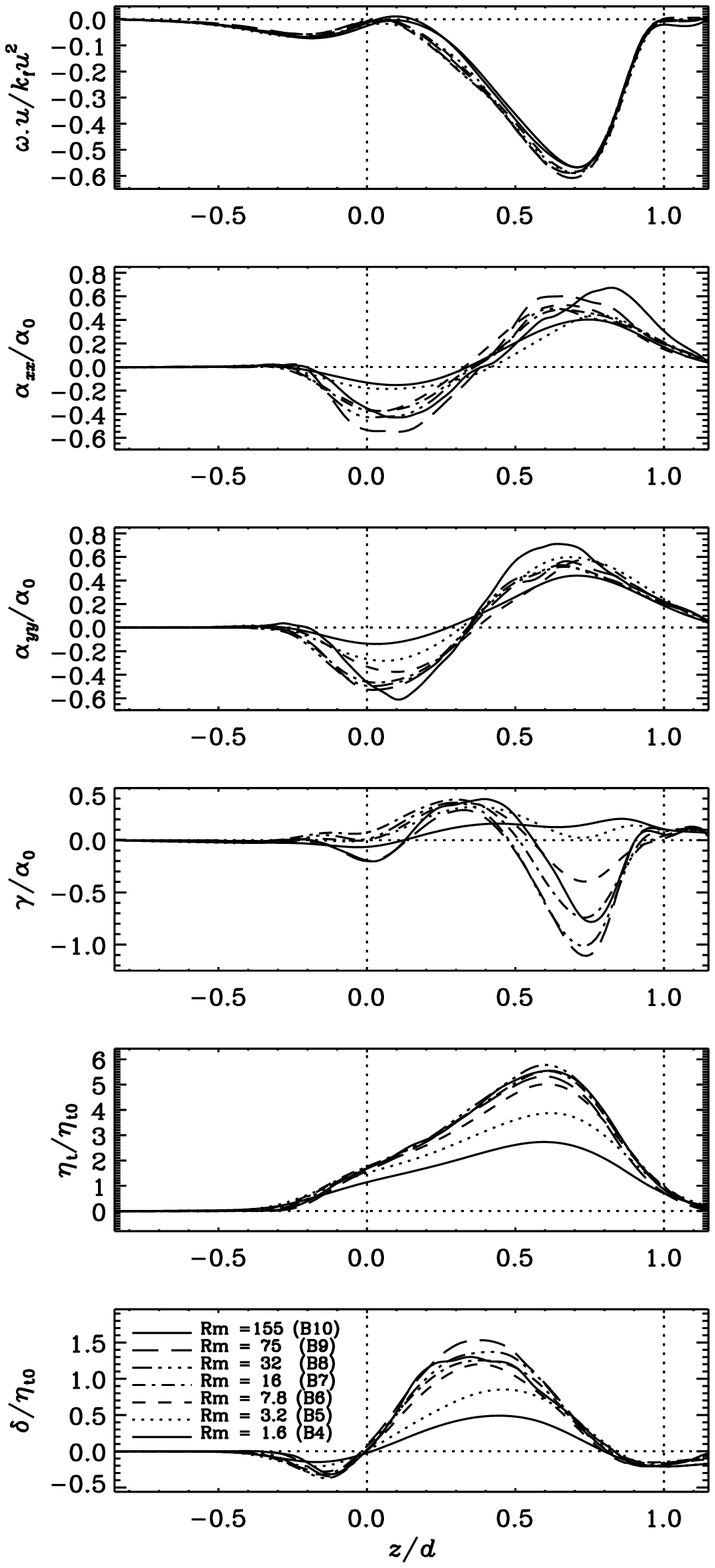}
\caption{From top to bottom: kinetic helicity, $\alpha_{xx}$,
  $\alpha_{yy}$, $\gamma$, $\etat$, and $\delta$ as functions of
  $\Rm$ from Runs~B4--B10. The linestyles are as indicated in the lowermost
  panel. $\Co\approx0.33$, $\theta=\Sh=0$ and $\Rey\approx15$ in all runs.}
\label{fig:palprm_cpm}
\end{figure}

\begin{figure}[t]
\centering
\includegraphics[width=\columnwidth]{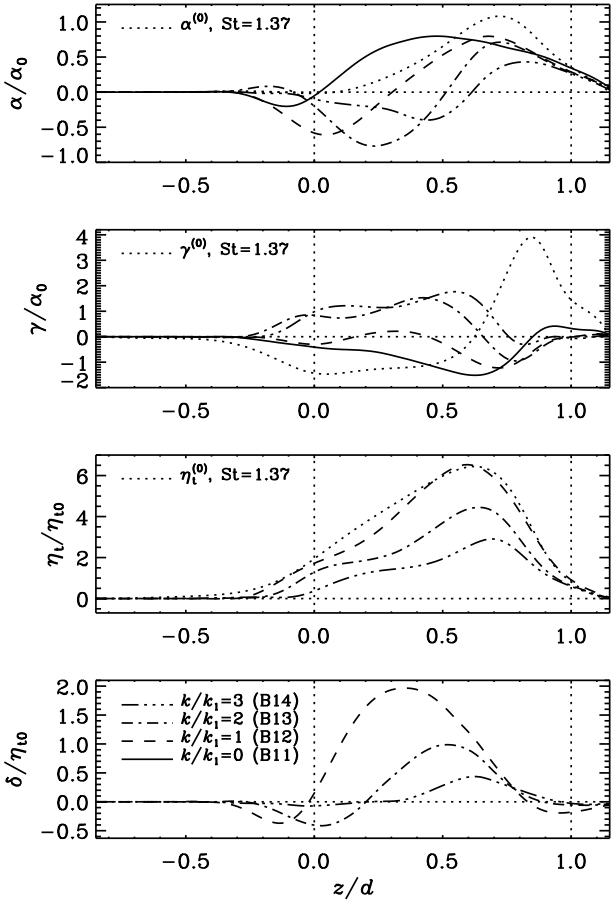}
\caption{From top to bottom: $\alpha$, $\gamma$, $\etat$, and $\delta$
  as functions of $k$ for runs B11--B14 with $\Rm\approx35$ and
  $\Co\approx0.36$. Linestyles as indicated in the lowermost panel.}
\label{fig:kdep_128b}
\end{figure}

\begin{figure}[t]
\centering
\includegraphics[width=\columnwidth]{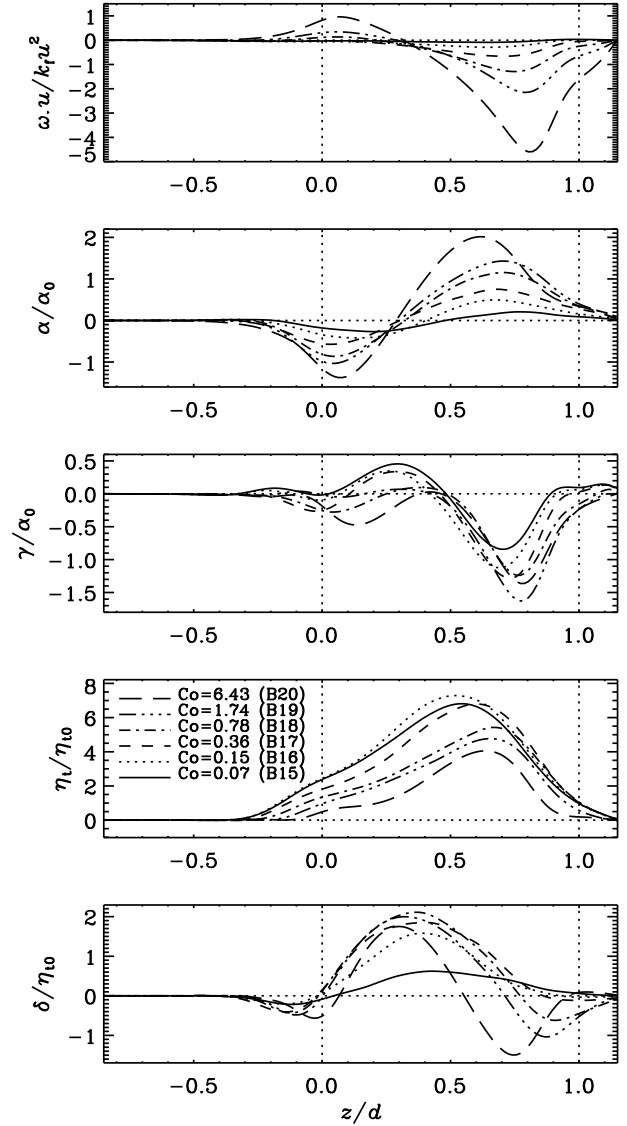}
\caption{From top to bottom: kinetic helicity, $\alpha$, $\gamma$,
  $\etat$, and $\delta$ as functions of rotation from Runs~B15--B20. The linestyles are
  as indicated in the second panel from the bottom. $\theta=0$, $\Sh=0$, and
  $\Rm\approx20\ldots35$.}
\label{fig:palprot}
\end{figure}

\begin{figure}[t]
\centering
\includegraphics[width=\columnwidth]{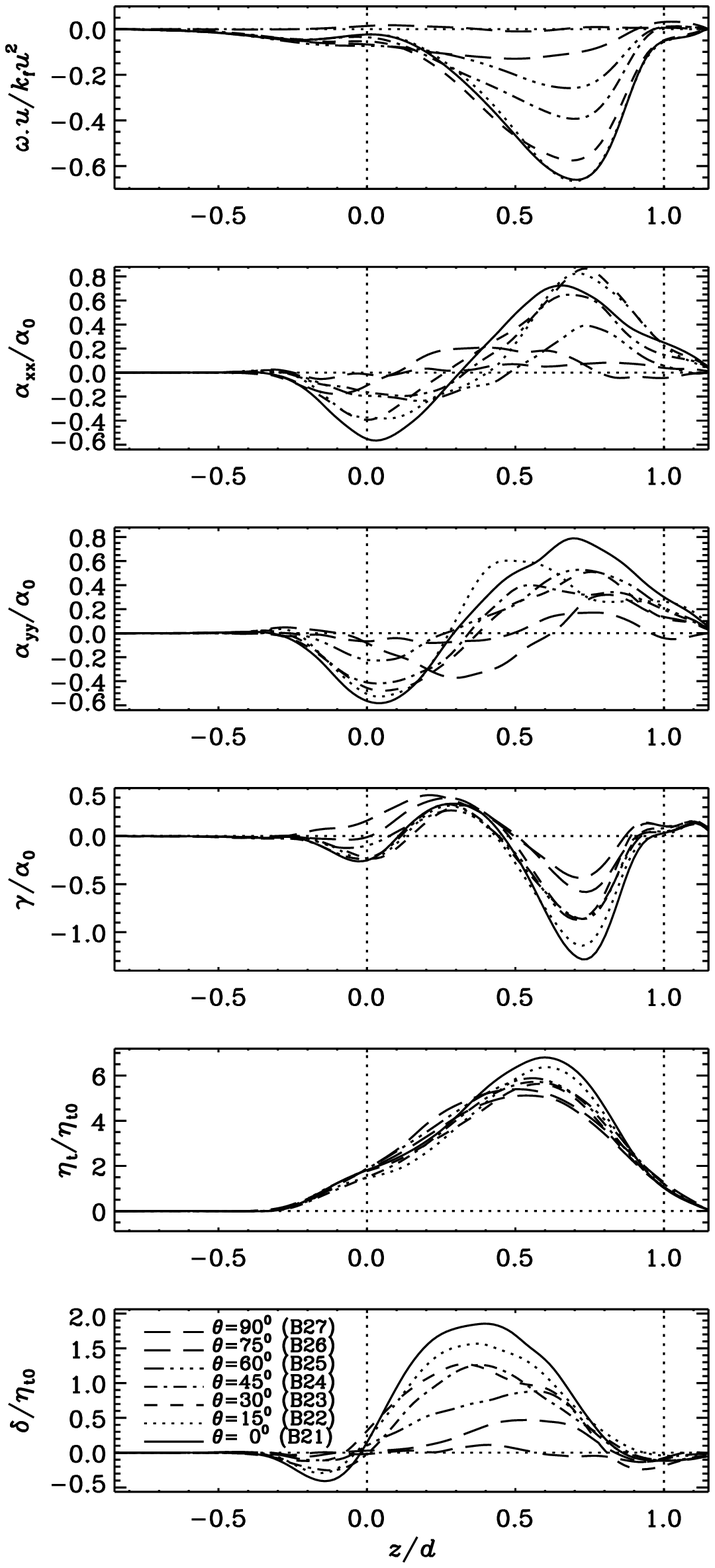}
\caption{From top to bottom: kinetic helicity, $\alpha_{xx}$,
  $\alpha_{yy}$, $\gamma$, $\etat$, and $\delta$ as functions of
  colatitude $\theta$ from Runs~B21--B27. The linestyles are as indicated in the
  lowermost panel. $\Co\approx0.29-0.36$ and $\Rm\approx35-43$ in all runs.}
\label{fig:palplat}
\end{figure}

\subsection{Set B: only rotation $(\Co\neq0$, $\Sh=0)$}
When rotation (corresponding to the north pole, $\theta=0$) is added to
the system, non-zero negative kinetic helicity is produced due to the
fact that $\bm{g}\cdot\bm{\Omega}<0$. Although the $\alpha$-effect is
not directly proportional to the helicity in the anisotropic case of
stratified convection, it can still be a useful proxy. 
Figure \ref{fig:aerot} shows the results for Run~B with
$\Co\approx0.36$ and $\Rm\approx35$.
We find that
the diagonal components of $\alpha_{ij}$ are positive in the upper
part of the convectively unstable region where the helicity is most
negative. The negative maxima at the base of the convection zone are,
however, not reflected by the kinetic helicity.
The results in Fig.~\ref{fig:aerot} were obtained for $k/k_1=1$.
The profiles of $\alpha_{xx}$ and $\alpha_{yy}$
are more in line with the profile of the helicity for $k/k_1=0$
(see Sect.~\ref{sec:kdep} for more details on the $k$-dependence).

In comparison to Run~A,
the pumping coefficient $\gamma$ shows a deeper maximum in the upper
half of the convection zone and somewhat decreased value in the lower
half. The profile and magnitude of the turbulent diffusivity are
similar to those in the nonrotating case. The
coefficients $\eta_{yx}$ and $\eta_{xy}$ are equal in magnitude and of
opposite sign. This leads to a positive (negative) $\delta$ in the
convection zone (overshoot layer) with magnitude peaking close to twice
$\etatz$. The quantities $\epsilon_\gamma$ and $\epsilon_\eta$
are small, as expected from symmetry arguments

\subsubsection{Dependence on horizontal system size}
The profiles and magnitudes of the two diagonal components of $\alpha$ are very
close to each other in our standard case (Run~B) shown in
Fig.~\ref{fig:aerot}. Although the number of convection cells in the
domain is quite small (of the order of ten or less), the isotropy of
$\alpha$ is a preliminary indication that the horizontal system is still large
enough to give representative results relevant for a larger ensemble.
If the system size is too small, the derived turbulent transport coefficients may
no longer be meaningful (Hughes \& Cattaneo \cite{HC08}).
In order to study
the convergence of our results, we have performed simulations with
three box sizes where the horizontal extent $L_{\rm H}\equiv L_x=L_y$ is
either $2d$, $4d$, or $8d$, respectively, and the vertical
extent of the box is kept unchanged. These runs are labeled (from the smallest 
to the largest) as B1, B2, and B3, where B2 is the same as Run~B (cf.\ Fig.~\ref{fig:aerot}).
All three runs are relatively slowly
rotating with $\Co\approx0.36$, $\Sh=0$, and $\Rm\approx35$.
The results are shown in Fig.~\ref{fig:palpsize}. It is obvious that
the differences between the runs are very small and the two larger
systems are virtually identical. The only statistically significant
difference is the anisotropy of $\alpha$ in the convectively
unstable region for Run~B1 with the smallest system size. We can thus be fairly
confident that the standard box size with $L_{\rm H}=4d$ is
sufficiently large.
This is consistent with the results of Hughes \& Cattaneo
(\cite{HC08}) who found that the ratio $L_{\rm H}/d$ needs to be
larger than two for the value of $\alpha$ to be reasonably
representative of the system.

\subsubsection{Dependence on $\Rm$}
One of the basic expected properties of turbulent dynamos is that they should be
`fast', i.e.\ the growth rate of the dynamo, and thus the
transport coefficients, should not depend on the molecular magnetic
diffusion provided that $\Rm\gg1$.
Figure \ref{fig:palprm_cpm} shows the transport coefficients as
functions of $\Rm$ for fixed $\Rey\approx15$, $\Co\approx0.33$, 
and $\theta=\Sh=0$ from Runs~B4--B10. The magnetic Reynolds
and Prandtl numbers vary in the ranges $1.6\ldots155$ and $0.1\ldots10$,
respectively.
We find that the transport coefficients show no statistically significant dependence
on $\Rm$ for $\Rm\ga8$. The only appreciable departures occur for the
two lowest Reynolds numbers.
However, our definition of the
Reynolds number depends on $\kef$ ($=2\pi/d$) which is not as well defined as in,
e.g., forced turbulence simulations,
so the coefficients may depend on $\eta$ for values somewhat larger than 
$\Rm=1$ in the present case.
These results agree with those
obtained for isotropic turbulence (Sur et al.\ \cite{SBS08};
Brandenburg et al.\ \cite{BRRK08}; Mitra et al.\ \cite{MKTB08}).

\subsubsection{Dependence on wavenumber $k$}
\label{sec:kdep}
The results for nonrotating convection (see Fig.~\ref{fig:kdep_128a})
indicate that at least the
pumping effect can experience not only a change in magnitude but also a
qualitative change when the wavenumber of the test field is
varied (for corresponding details see Brandenburg et al.\ \cite{BRS08}).
It is of great interest to study whether similar effects can
occur for the $\alpha$-effect. Our results for the standard case of
$\Rm\approx35$ and $\Co\approx0.36$ from runs B11--B14 are shown in
Fig.~\ref{fig:kdep_128a}. 
According to symmetry arguments, the diagonal components of
$\alpha_{ij}$ are the same for $\theta=0$. We confirm this
numerically (see the two previous sections) and combine the data of the
$\alpha$-effect into a single coefficient $\alpha$.
The same applies to the FOSA expressions,
Eqs.~(\ref{equ:alpxx})--(\ref{equ:alpyy}), which we combine into
$\alpha^{(0)}=\onehalf(\alpha^{(0)}_{xx}+\alpha^{(0)}_{yy})$.
For $k/k_1=0$, the $\alpha$-effect shows a
more uniform positive value in the convectively unstable region than
for $k/k_1=1$. For larger $k$ the negative region in the deeper layers
shifts towards the top and the maxima of the profile diminish. The
pumping coefficient and turbulent diffusion behave very much the same
as in the nonrotating case, cf.\ Fig.~\ref{fig:kdep_128a}. The coefficient
$\delta$ diminishes rapidly as $k$ increases.
The FOSA expression for the $\alpha$-effect is in qualitative
accordance with the $k/k_1=0$ result, but fails to capture the details
of the profile (see also K\"apyl\"a et al.\ \cite{KKOS06}). 
The Strouhal number required to match $\etat^{(0)}$
with $\etat$ is 1.37 which is somewhat smaller than in the nonrotating
case.

\subsubsection{Dependence on $\Co$}
The $\alpha$-effect from rotating convection simulations in setups 
similar to ours has been
studied in numerous papers in the past using the imposed field method
(e.g.\ Ossendrijver et al.\ \cite{OSB01,OSRB02}; K\"apyl\"a et
al.\ \cite{KKOS06}). Our results for the Coriolis number dependence of
the kinetic helicity, $\alpha$, $\gamma$,
$\etat$ and $\delta$ from Runs~B15--B20 are shown in Fig.~\ref{fig:palprot}.
We find that the kinetic helicity and $\alpha$ increase monotonically as 
functions of rotation
in accordance with the results of Ossendrijver et al.\
(\cite{OSB01}). The vertical pumping effect shows little
dependence on rotation although the Coriolis number changes by two
orders of magnitude. 

Interestingly, the turbulent diffusivity shows a
marked decrease for rapid rotation. 
The coefficient $\delta$ is positive in the convection zone and
negative in the overshoot layer for slow rotation. The magnitude increases
rapidly until $\Co\approx0.2$, after which $\delta$ changes sign near
the top. This negative region increases with rotation. Similar
results, i.e.\ monotonically decreasing $\etat$ and a first increasing
and then decreasing $\delta$ were obtained from forced turbulence
simulations by Brandenburg et al.\ (\cite{BRRK08}).

The combined effect of increasing $\alpha$ and decreasing $\etat$
suggests that the large-scale dynamo was possibly subcritical in the
runs with only rotation in Paper I and other earlier studies (e.g.\ 
Nordlund et al.\ \cite{NBJRRST92}; Brandenburg et al.\ \cite{BJNRST96}; 
Cattaneo \& Hughes \cite{CH06}; Tobias et al.\ \cite{TCB08}), 
but that it could be excited for
more rapid rotation. The validity of this conjecture is given some
credibility by K\"apyl\"a et al.\ (\cite{KKB08b}) who find clear
large-scale dynamo action for $\Co>4$ for a similar setup as used here
and in Paper I. More detailed discussion of these results can be found
in the aforementioned reference.

\subsubsection{Dependence on $\theta$}
The latitude dependence of the coefficients for $\Co\approx0.29-0.36$ 
and $\Rm\approx35-43$ from Runs~B21--B27 is shown in 
Fig.~\ref{fig:palplat}. The colatitude is
varied from $0\degr$ (north pole) to $90\degr$ (equator) in increments of 15
degrees. The kinetic helicity and the diagonal components of $\alpha_{ij}$ 
decrease
monotonically towards the equator.
The latter are approximately equal and show
a similar latitude dependence.
This is consistent with the results
of Ossendrijver et al.\ (\cite{OSRB02}) and K\"apyl\"a et al.\
(\cite{KKOS06}) with a comparable Coriolis number\footnote{The
  definition of Coriolis number in the present study is smaller by a factor of
  $2\pi$ in comparison to previous studies.}.  Vertical pumping also
decreases monotonically from the pole to the equator. The equatorial
profile of $\gamma$ is quite similar to the nonrotating run, cf.\
Fig.~\ref{fig:aenorot}. The trend is similar to that seen in earlier
studies, e.g.\ K\"apyl\"a et al.\ (\cite{KKOS06}). The variation of
$\etat$ is smaller than for the other components, but a weak increasing
trend from the pole towards the equator is seen.
The coefficient $\delta$ shows a clear decreasing trend as a function
of colatitude and is consistent with zero at the equator --
in accordance with symmetry considerations.

\begin{figure}[t]
\centering
\includegraphics[width=\columnwidth]{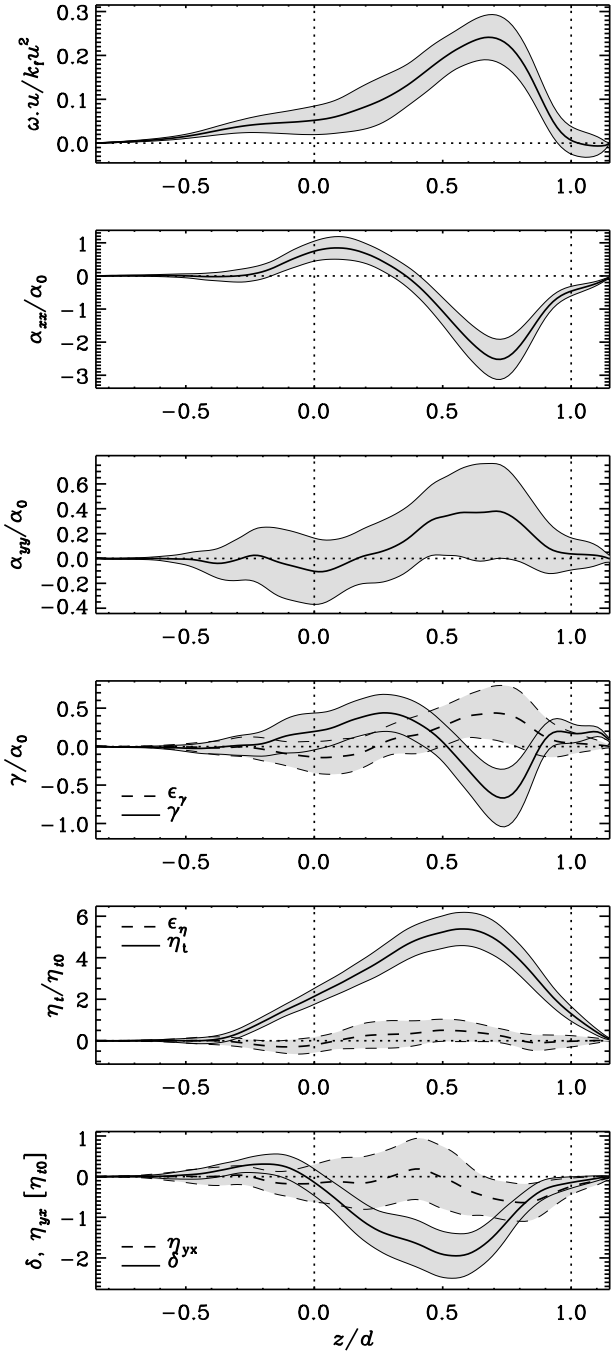}
\caption{Same as Fig.~\ref{fig:aenorot}, but for Run~C with no rotation
  and just shear; $\Co=0$, $\Sh\approx-0.14$, and $\Rm\approx46$.}
\label{fig:aeshe}
\end{figure}

\begin{figure}[t]
\centering
\includegraphics[width=\columnwidth]{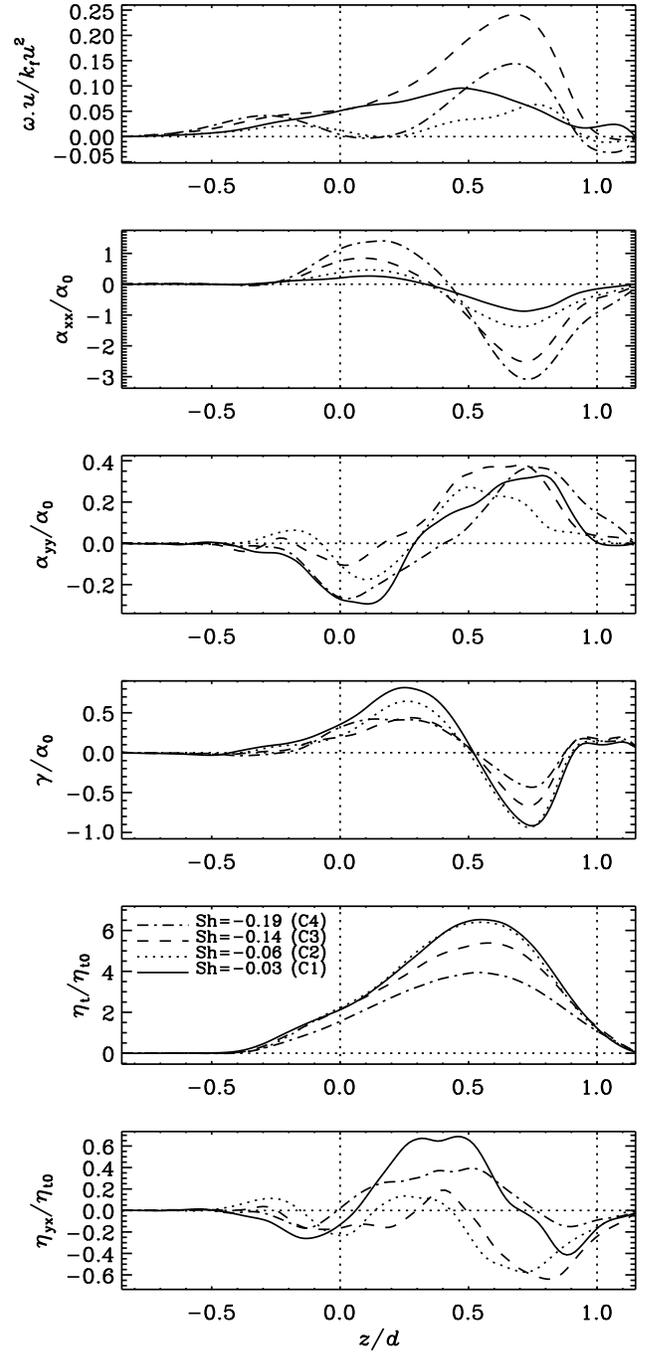}
\caption{From top to bottom: kinetic helicity, $\alpha_{xx}$,
  $\alpha_{yy}$, $\gamma$, $\etat$, and $\eta_{yx}$ as functions of shear
  from Runs~C1--C4. The
  linestyles are as indicated in the second panel from the below. $\Co=0$ and
  $\Rm\approx42-46$ in all runs.}
\label{fig:palpshe}
\end{figure}

\begin{figure}[t]
\centering
\includegraphics[width=\columnwidth]{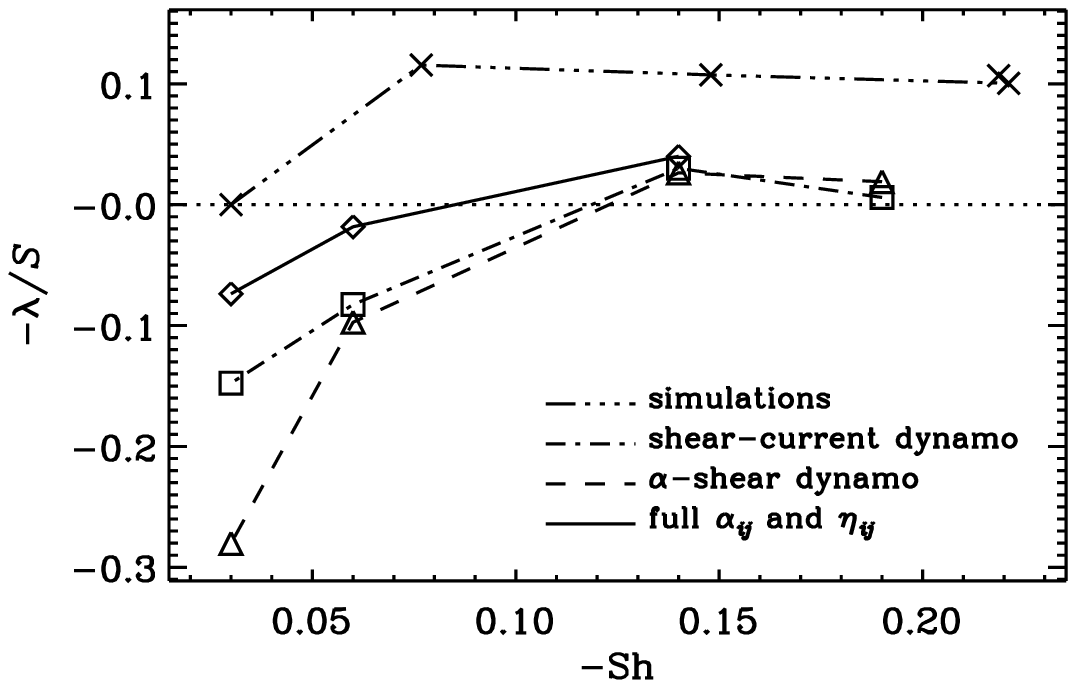}
\caption{Growth rates $\lambda$ from a one-dimensional
  mean-field model with transport coefficients from Runs~C1--C4
  presented in Fig.~\ref{fig:palpshe}. Linestyles as indicated in the
  lower panel.}
\label{fig:pgr_she}
\end{figure}

\subsection{Set C: only shear $(\Co=0$, $\Sh\neq0)$}

\subsubsection{Simulation results}
The next case to consider is that of shear only. We use uniform shear
of the form $\meanv{U}=(0,Sx,0)$, where $S<0$, 
resulting in $\bm{g}\cdot(\bm{\nabla}\times\meanv{U})>0$, and positive
kinetic helicity, as expected.
See Fig.~\ref{fig:aeshe} for representative results from Run~C with
$\Sh\approx-0.14$ and $\Rm\approx46$.
We find that introducing shear into the system produces an anisotropic
$\alpha$-effect.
The profile of the $\alpha_{xx}$ component is similar, but of opposite sign,
to that in the case with only rotation; see, e.g., Fig.~\ref{fig:aerot}. 
The magnitude of this
component is also quite large, i.e.\ up to twice the estimate
$\alpha_0$
already for rather weak shear of $\Sh\approx-0.14$.
The $\alpha_{yy}$ coefficient, relevant for dynamo excitation,
is positive, but
the magnitude is only about one fifth of $\alpha_{xx}$.
Moreover, the error bars are so large that the value is hardly statistically
significant. These
results demonstrate that linking the $\alpha$-effect to the negative
of the kinetic helicity can be misleading.
Figure \ref{fig:palpshe} shows the coefficients as functions of shear
for Runs~C1--C4.
We have to restrict the study to rather modest values of $\Sh$ because
shear, in the absence of rotation, promotes generation of large-scale vorticity
(e.g.\ Elperin et al.\ \cite{EKR03}; K\"apyl\"a et al.\
\cite{KMB08}). Although our largest value of $\Sh$ is still rather modest,
the $\alpha_{xx}$ component is quite large, up to three times
$\alpha_0$. The $\alpha_{yy}$ component, however, remains small and
positive for all values of $\Sh$ without a
consistent trend as a function of shear.

The turbulent pumping in Run~C has a similar profile as in the cases with
$\Co=\Sh=0$ (Run~A) and $\Co\neq0$ (Run~B) with $k/k_1=1$ with 
downward pumping near the surface and
upward pumping in the lower part of the convectively unstable region.
The profile and magnitude of the turbulent diffusivity is also very
similar to previous cases.
The pumping effect and turbulent diffusivity are
decreased when the magnitude of $\Sh$ is greater than 0.06. The
results for increasing $\alpha$ and decreasing $\etat$ as functions of
shear are opposite to those obtained from helically forced turbulence
with shear (Mitra et al.\ \cite{MKTB08}).
However, the comparison for the $\alpha$-effect should be done with
caution because in Mitra et al.\ (\cite{MKTB08}) $\alpha$ arises
essentially due to the external forcing and is only modified by the
action of shear whereas in the present case $\alpha$ is due to the
interaction of shear, stratification, and turbulence themselves.

The $\eta_{yx}$ component, which can drive a mean-field shear-current
dynamo for $\eta_{yx}S > 0$, is of interest because it can
provide an explanation for the dynamos seen in recent dynamo
simulations (Paper I; Hughes \& Proctor \cite{HP08}).  In the present
case where $S<0$, $\eta_{yx}$ should be negative to excite the
shear-current dynamo. There appear to be consistently negative regions
of $\eta_{yx}$ at the interface of the convectively unstable region
and the overshoot layer, and in the upper layers of the convection
zone.  The upper negative region is more pronounced for $\Sh=-0.06$
and $\Sh=-0.14$. However, the errors of these quantities are of the
same order of magnitude as the mean value, cf.\ the bottom panel of
Fig.~\ref{fig:aeshe}. These results tend to agree with earlier
findings from forced turbulence (Brandenburg et al.\ \cite{BRRK08};
Mitra et al.\ \cite{MKTB08})
where $\eta_{yx}$ for the most part was positive or compatible with
zero.

\subsubsection{Mean--field dynamo models}
\label{sec:mfshe}
In Paper I clear large-scale dynamo action was found from a simulation
with $\Sh\approx-0.08$ whereas for $\Sh\approx-0.03$ the solution was
marginal. In the range $-0.08 > \Sh > -0.22$ the growth rate of the
large-scale field was proportional to the modulus of the shear parameter $S$.
Bearing these results in mind and having obtained the turbulent transport
coefficients for the corresponding parameter regime we are in a
position to apply the test field results in a mean-field dynamo model.

Using the full test field results for $\alpha_{ij}$ and $\eta_{ij}$ in the
corresponding mean-field model indicates that a dynamo is excited for $-\Sh>0.14$,
see the growth rates of the large-scale field presented Fig.~\ref{fig:pgr_she}.
It is interesting to study what are the relative importances of the
different effects: first we turn off the off-diagonal components of
$\eta_{ij}$ in which case the magnetic field is generated by the
$\alpha$-effect and the shear-current effect is absent. We find that
the growth rate decreases but is still positive for the same cases as
before. On the other hand, a ``pure'' shear-current dynamo, i.e.\ where 
$\alpha_{ij}=\eta_{xy}=0$, is also excited
for the same runs with a very similar $\lambda$ as in the
$\alpha$-shear case. In comparison to the simulations of Paper I, we
find that the growth rates from the mean-field model are consistently
significantly smaller. These results and the fact that no dynamo was
found for $-\Sh=0.06$ would seem to indicate that an incoherent
$\alpha$-shear dynamo (e.g.\ Vishniac \& Brandenburg \cite{VB97}) is
also operating in the full simulations. However, we should remain
cautious when comparing the direct simulations and the mean-field model
because the transport coefficients were determined for a single value
of $k$ whereas many other wavenumbers are available in the
simulations.

\begin{figure}[t]
\centering
\includegraphics[width=\columnwidth]{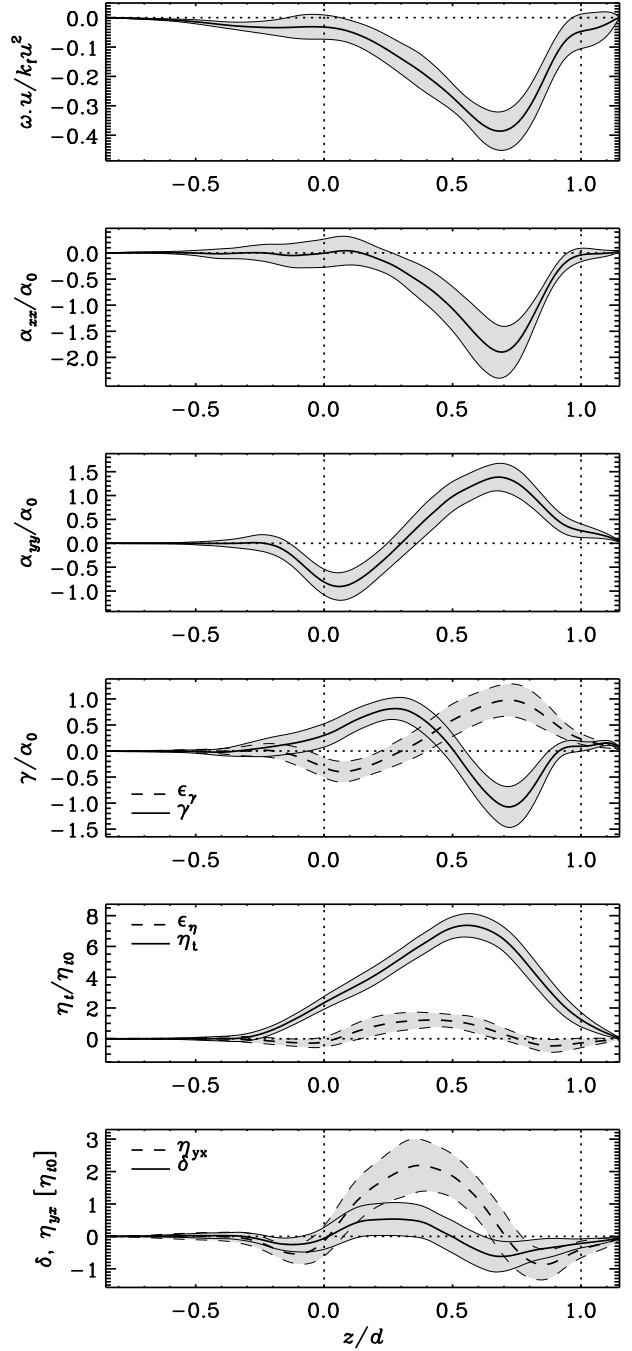}
\caption{Same as Fig.~\ref{fig:aenorot}, but for Run~D with both rotation
  and shear; $\Co\approx0.36$, $\Sh\approx-0.18$, and $\Rm\approx37$.}
\label{fig:aesherot}
\end{figure}

\begin{figure}[t]
\centering
\includegraphics[width=\columnwidth]{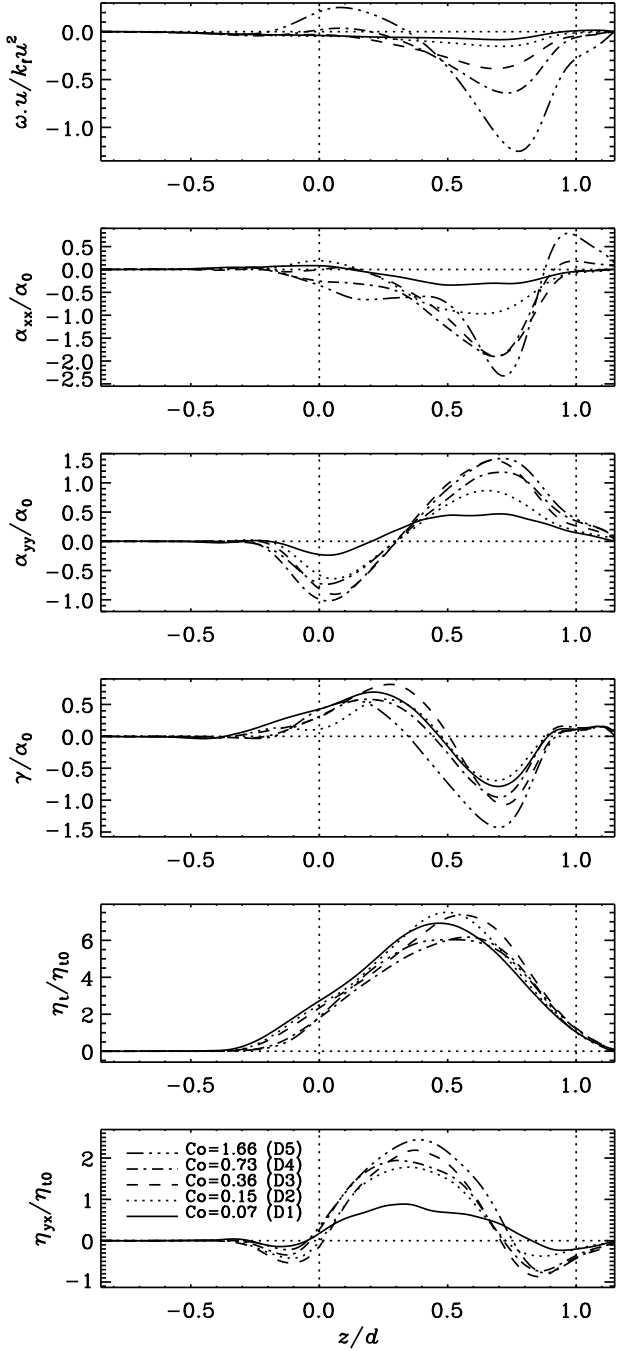}
\caption{From top to bottom: kinetic helicity, $\alpha_{xx}$,
  $\alpha_{yy}$, $\gamma$, $\etat$, and $\eta_{yx}$ as functions of Coriolis
  number for Runs~D1--D5. The linestyles are as indicated in the lowermost
  panel. $\Sh=-\onehalf \Co$ and $\Rm\approx37$ in all runs.}
\label{fig:alpsherot}
\end{figure}

\begin{figure}[t]
\centering
\includegraphics[width=\columnwidth]{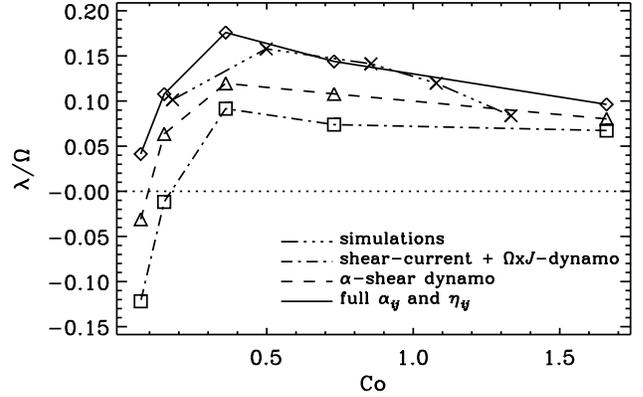}
\caption{Same as Fig.~\ref{fig:pgr_she} but for Runs~D1--D5 shown 
in Fig.~\ref{fig:alpsherot}.}
\label{fig:pgr_sherot}
\end{figure}

\subsection{Set D: rotation and shear $(\Co\neq0$, $\Sh\neq0)$}
\subsubsection{Simulation results}
When rotation is added to the system where a large-scale shear is
already imposed, the vorticity generation is suppressed (e.g.\ Yousef et al.\ \cite{YHSea08b}; Paper I) and it is
possible to study higher values of $\Sh$. Representative results for
Run~D with $\Co\approx0.36$, $\Sh\approx-0.18$, and $\Rm\approx35$ are
shown in Fig.~\ref{fig:aesherot}. The results for the kinetic helicity
and $\alpha_{yy}$ seem to behave additively when comparing with the
runs with only rotation (Run~B; Fig.~\ref{fig:aerot}) and only shear
(Run~C; Fig.~\ref{fig:aeshe}). The $\alpha_{xx}$ component is somewhat
smaller than in Run~C with shear only
which is consistent with oppositely signed contributions due to shear
and rotation.

We note that in a recent paper, Hughes \& Proctor (\cite{HP08}) found that the
$\alpha$-effect is virtually unchanged when shear is added to a
rotating system. In their case the shear profile is proportional to
$\cos y$. The resulting large-scale vorticity is then $\mean{W}_z
\propto \sin y$ which leads to $\alpha^{(W)}_{ij} \propto
(\bm{G}\cdot\meanv{W})\delta_{ij} \propto \sin y$ (e.g.\ R\"adler \&
Stepanov \cite{RS06}) where $\bm{G}$ symbolically denotes the
inhomogeneity of the turbulence. However, Hughes \& Proctor
(\cite{HP08}) show a volume average of $\alpha$ over the full upper half
of the domain in which case the contribution of $\alpha^{(W)}$ cancels out.
This explains the absence of any modifications of $\alpha$
due to shear in their case, but for us this is not the case because for
our shear profile $\mean{W}_z=S=\mbox{const}$.

The profile of $\gamma$ is quite similar to the rotating case, i.e.\
Run~B; see Fig.~\ref{fig:aerot}, with the exception that the
off-diagonal components of $\alpha_{ij}$ show considerable anisotropy
as manifested by the parameter $\epsilon_\gamma$. The turbulent
diffusion shows a profile common to all the other simulations, but
here the diagonal components of $\eta_{ij}$ show evidence of mild anisotropy with
$\epsilon_\eta$ peaking near the middle of the convectively unstable
region with a maximum value of $\epsilon_\eta \approx \etatz$. The
quantity $\delta$ is compatible with zero whereas $\eta_{yx}$ exhibits
a similar profile and magnitude as $\delta$ does in the rotating
case, cf.\ Fig.~\ref{fig:aerot}, which indicates that
$\eta_{xy}\approx\eta_{yx}$, as opposed to $\eta_{xy}\approx-\eta_{yx}$
in Run~B.

The kinetic helicity and the turbulent transport coefficients as functions of
$\Co$, keeping the ratio $-S/\Omega=1$ constant, are shown in
Fig.~\ref{fig:alpsherot}. The helicity is increasing in a similar
fashion as, albeit slower than, in the absence of shear, compare with
Fig.~\ref{fig:palprot}. The components of the $\alpha$-effect are
highly anisotropic with the main contribution of $\alpha_{xx}$ being
due to shear and that of $\alpha_{yy}$ due to rotation (compare with 
Figs.~\ref{fig:palprot} and \ref{fig:palpshe}). The value of
$\alpha_{xx}$ is somewhat decreased in comparison to the cases with
shear only whereas $\alpha_{yy}$ is almost unaffected. This is in
qualitative agreement with adding the contributions of runs
from Sets~B and C with corresponding $\Co$ and $\Sh$, respectively.

The profiles of $\gamma$ and $\etat$ are similar to those
with only
rotation. The differences of $\gamma$ and $\etat$ as a function of
$\Co$ are small and for the most part fall within the error bars.
The profile of $\eta_{yx}$ is very similar to $\delta$ in the case of only
rotation with negative values near the base and top of the
convectively unstable region with positive values in between.
The profile and magnitude of $\eta_{yx}$ remains essentially fixed for
$\Co\ga0.15$.
We note that in the simulations with shear and rotation the
$\bm{\Omega}\times{\bm{J}}$-effect may also contribute to the
generation of large-scale magnetic fields (e.g.\ R\"adler \cite{R69};
R\"adler et al.\ \cite{RKR03}; Pipin et al.\ \cite{P08}). It is,
however, not altogether clear how to disentangle the transport
coefficients responsible for the shear-current and
$\bm{\Omega}\times{\bm{J}}$-dynamos in the present case. Thus the
off-diagonal components of $\eta_{ij}$ contain contributions from
both effects in Set~D.

\subsubsection{Mean--field dynamo models}
We follow here the same procedure as in Sect.~\ref{sec:mfshe} to study 
dynamo excitation for the Runs~D1--D5.
Using the full $\alpha_{ij}$ and $\eta_{ij}$ tensors in the
one-dimensional mean-field model indicates that all of the runs in
Fig.~\ref{fig:alpsherot} are capable of driving a dynamo. Neglecting
the off-diagonal components of $\eta_{ij}$ decreases $\lambda$ by
approximately a third whereas for a pure shear-current dynamo the
growth rate is roughly half of the model where the full $\alpha_{ij}$
and $\eta_{ij}$ tensors were used. For an $\alpha$-shear dynamo
$\lambda$ is positive for all runs except the slowest rotation case
with $\Co=0.07$ whereas for the combined shear-current and 
$\bm{\Omega}\times{\bm{J}}$-dynamo also $\Co=0.15$
is mildly subcritical. It is interesting to note that the simulation results
of Paper I fall roughly on top of the uppermost line, i.e.\ where the
full $\alpha_{ij}$ and $\eta_{ij}$ are used, in
Fig.~\ref{fig:pgr_sherot}. This is in contrast to the case of only
shear where the growth rates from the mean-field model were clearly
smaller than in the corresponding simulations of Paper I.
However, the mean-field model does not reproduce the declining growth
rate for $\Co\ga1.1$ that is observed in the direct
simulations.

\section{Conclusions}
We obtain turbulent transport coefficients governing the evolution of
large-scale magnetic fields from turbulent convection simulations with
the test field method. 
We study the system size and magnetic Reynolds number dependences of
the coefficients. This is important because spurious results can be
expected for small Reynolds numbers or when the aspect ratio of the domain
is too small (Hughes \& Cattaneo \cite{HC08}).
We find that for our standard system size, $L_{\rm H}/d=4$,
the coefficients are essentially identical to those obtained
with a horizontal extent that is twice as large. As a function of $\Rm$, all
the coefficients are essentially constant for $\Rm\ga8$.
This is in accordance with the theory but at odds with results
from certain imposed field calculations (e.g.\ Cattaneo \& Hughes
\cite{CH06}).
In these calculations the magnetic field is allowed to evolve until
saturation which can cause strong quenching even if the imposed field
itself is weak. This is particularly important if closed boundary
conditions for the magnetic field are imposed, in which case magnetic
helicity conservation can lead to catastrophic quenching 
(Vainshtein \& Cattaneo \cite{VC92}). More reliable
results for the kinematic $\alpha$-effect
with the imposed field method can be obtained by resetting the
magnetic field before it grows too large or develops substantial
gradients (e.g.\ Ossendrijver at al.\ \cite{OSRB02}; K\"apyl\"a et
al.\ \cite{KKOS06}). More detailed comparison of the imposed field and
test field methods is important, but beyond the scope of the present
study.

The earlier determinations of transport coefficients from convection
simulations have used the imposed field method (e.g.\ Ossendrijver et
al.\ \cite{OSB01,OSRB02}; K\"apyl\"a et al.\ \cite{KKOS06}) which
yields the components of $\alpha_{ij}$ but does not deliver
$\eta_{ij}$ because the imposed field is uniform. The test field
method does not suffer from this restriction and $\eta_{ij}$ and the
$k$-dependence of the coefficients can be extracted. We find that for
$k/k_1=0$, i.e.\ for a uniform field, the results for $\alpha$ and
$\gamma$ are consistent with those obtained from imposed field
calculations, provided the magnetic field is reset before it grows
too large and substantial gradients develop.
As $k$ is increased, however, the qualitative behaviour
of the coefficients changes. This is indicated by a partial
sign change of $\alpha$ and a complete sign change of $\gamma$; see
Figs.~\ref{fig:kdep_128a} and \ref{fig:kdep_128b}.

The turbulent diffusivity shows a robust behaviour regardless of
the parameters of the simulations: the profile is proportional to the
vertical velocity squared, $\mean{u_z^2}$, as predicted by FOSA (e.g.\
R\"adler \cite{R80}). The value of $\etat$ decreases almost
proportional to $k^{-1}$, and shows a declining trend as a function
of rotation and shear.

For the present parameters,
the $\alpha$-effect increases monotonically as rotation is increased.
As a function of latitude, the diagonal components of $\alpha_{ij}$
have a similar magnitude and peak near the pole with declining values
towards the equator. The $\alpha$-effect induced by shear is highly
anisotropic: the $\alpha_{xx}$ component has a similar profile and
magnitude, but opposite sign, as $\alpha_{xx}$ and $\alpha_{yy}$ in the case of only
rotation. This component also increases monotonically as a function of
shear, whereas the shear-induced $\alpha_{yy}$ remains small
regardless of the strength of the shear. In the runs where rotation
and shear are present, the diagonal components of $\alpha_{ij}$ are
roughly the sums of the corresponding coefficients in the cases with
rotation and shear alone.

In addition to the $\alpha$-effect, the $\eta_{yx}$ component can contribute
to a shear-current dynamo when $\eta_{yx}S>0$. In our case, where $S<0$,
such dynamo action is possible if $\eta_{yx}<0$. We find that this
coefficient shows negative regions near the base and near the top of the
convectively unstable region, but the errors are of the same order 
of magnitude as the negative mean values in most cases. 

In order to connect to earlier work, we use the test field results in
a one-dimensional mean-field model in order to understand the
excitation of dynamos using identical setups as in direct simulations
(Paper I). We study here only the cases with shear and
consider large-scale dynamos in the rigidly rotating case elsewhere
(K\"apyl\"a et al.\ \cite{KKB08b}).
The presently used dynamo model ignores $k$-dependence and is therefore
likely to be too simple to fully describe the large-scale fields in the
direct simulations. Nevertheless, the present results, taken at face value,
seem to indicate that in the case with shear alone the derived dynamo
coefficients are not sufficient to explain the dynamo but that an
additional incoherent $\alpha$-shear dynamo might be needed. This
conjecture is based on the fact that mean-field $\alpha$-shear and
shear-current dynamos are both excited with similar growth rates
which, however, are significantly smaller than those obtained from
direct simulations in Paper I.
Furthermore, a large-scale dynamo was marginal for $\Sh=-0.03$ in
Paper I, whereas for $\Sh=-0.06$ it was found to be slightly subcritical in the
present study.
On the other hand, for the case with both shear and rotation, no
additional incoherent effects seem to be needed. We find that in this
case the regular $\alpha$-shear dynamo produces larger growth rates
than the combined shear-current and $\bm{\Omega}\times\bm{J}$ dynamo but neither effect alone seems to be
strong enough to explain the dynamos in Paper I.

On a more general level, mean-field dynamo models of the Sun and other
stars rely on parameterisations of turbulent transport
coefficients. Even today, the majority of solar dynamo models bypass
this problem and ignore most of the turbulent effects and rely on
phenomenological descriptions of the $\alpha$-effect and turbulent
diffusion that are not without problems theoretically. On the other
hand, some attempts have been made to incorporate the results for the
transport coefficients from imposed field studies in mean-field models
of the solar magnetism (e.g.\ K\"apyl\"a et al.\ \cite{KKT06}; 
Guerrero \& de Gouveia Dal Pino \cite{GG08}) and
models employing more general turbulence models have recently appeared
(e.g.\ Pipin \& Seehafer \cite{PS08}). We feel that this is a worthy cause to
follow further with the present results.

\begin{acknowledgements}
  The authors wish to acknowledge the anonymous referee and Prof.\ 
  Gunther R\"udiger for their helpful comments on the manuscript.
  The computations were performed on the facilities hosted by
  CSC -- IT Center for Science in Espoo, Finland, who are administered
  by the Finnish ministry of education.
  This research has greatly benefitted 
  from the computational resources granted by the CSC to the grand 
  challenge project `Dynamo08'. 
  Financial support from the Academy of Finland grants No.\ 121431 (PJK)
  and 112020 (MJK) and the Swedish Research Council grant 621-2007-4064 (AB)
  is acknowledged.
  The authors acknowledge the hospitality of Nordita during the program
  `Turbulence and Dynamos' during which this work was initiated.
\end{acknowledgements}



\begin{thebibliography}{}

\bibitem[2001]{B01} Brandenburg, A. 2001, \apj, 550, 824

\bibitem[2005a]{B05a} Brandenburg, A. 2005a, \apj, 625, 539

\bibitem[2005b]{B05b} Brandenburg, A. 2005b, AN, 326, 787

\bibitem[2008]{B08} Brandenburg, A. 2008, AN, 329, 725

\bibitem[2007]{BK07} Brandenburg, A., \& K\"apyl\"a, P. J. 2007,
  NJP, 9, 305

\bibitem[2005]{BS05} Brandenburg, A., \& Subramanian, K. 2005,
  Phys.\ Rep.\ 417, 1

\bibitem[1990]{BTNPS90} Brandenburg, A., Tuominen, I., Nordlund,
  \AA., et al. 1990, \aap, 232, 277

\bibitem[1995]{BNST95} Brandenburg, A., Nordlund, \AA., Stein,
  R. F. \& Torkelsson U. 1995, \apj, 446,741

\bibitem[1996]{BJNRST96} Brandenburg, A., Jennings, R. L.,
  Nordlund, \AA., et al. 1996, JFM, 306, 325

\bibitem[2001]{BBS01} Brandenburg, A., Bigazzi, A. \& Subramanian,
  K. 2001, MNRAS, 325, 685

\bibitem[2008a]{BRRK08} Brandenburg, A., R\"adler, K.-H., Rheinhardt,
  M., \& K\"apyl\"a, P. J. 2008a, \apj, 676, 740

\bibitem[2008b]{BRS08} Brandenburg, A., R\"adler, K.-H. \& Schrinner, M.
  2008b, \aap, 482, 739

\bibitem[2007]{BBBMNT07} Brown, B. P., Browning, M. K., Brun, A. S.,
  Miesch, M. S., et al. 2007, AIPC, 948, 271

\bibitem[2006]{BMBT06} Browning, M. K., Miesch, M. S., Brun, A. S. \&
  Toomre, J. 2006, \apj, 648, L157
 
\bibitem[2006]{CH06} Cattaneo, F. \& Hughes, D. W. 2006, JFM, 553,
   401

 \bibitem[2003]{EKR03} Elperin, T., Kleeorin, N. \& Rogachevskii,
   I. 2003, PhRvE, 68, 016311

 \bibitem[2005]{GZR05} Giesecke, A., Ziegler, U. \& R\"udiger,
   G. 2005, Phys. Earth Planet. Interiors, 152, 90

\bibitem[2008]{GZER08} Gressel, O., Ziegler, U., Elstner, D. \&
   R\"udiger, G. 2008, AN, 329, 619 401

 \bibitem[2008]{GG08} Guerrero, G. \& de Gouveia Dal Pino, E. M. 2008,
   \aap, 485, 267

 \bibitem[2008]{HC08} Hughes, D. W. \& Cattaneo, F. 2008, JFM, 594,
   445

 \bibitem[2009]{HP08} Hughes, D. W. \& Proctor, M. R. E. 2009,
   PhRvL, 102, 044501

 \bibitem[1996]{HGB96} Hawley, J. F., Gammie, C. F. \& Balbus,
   S.A. 1996, \apj, 464, 690

\bibitem[2009]{KB08} K\"apyl\"a, P. J., \& Brandenburg, A. 2009,
  \apj, in press, arXiv:0810.2298

\bibitem[2004]{KKT04} K\"apyl\"a, P. J., Korpi, M. J. \& Tuominen,
  I. 2004, \aap, 422, 793

\bibitem[2006a]{KKOS06} K\"apyl\"a, P. J., Korpi, M. J.,
  Ossendrijver, M. \& Stix, M. 2006a, \aap, 455, 401

\bibitem[2006b]{KKT06} K\"apyl\"a, P. J., Korpi, M. J.,
  \& Tuominen, I. 2006b, AN, 327, 884

\bibitem[2008]{KKB08} K\"apyl\"a, P. J., Korpi, M. J., \&
  Brandenburg, A. 2008, \aap, 491, 353 (Paper I)

\bibitem[2009a]{KKB08b} K\"apyl\"a, P. J., Korpi, M. J., \&
  Brandenburg, A. 2009a, \apj, 697, 1153

\bibitem[2009b]{KMB08} K\"apyl\"a, P. J., Mitra, D., \&
  Brandenburg, A. 2009b, PhRvE, 79, 016302

\bibitem[2008]{KR08} Kleeorin, N. \& Rogachevskii, I. 2008, PhRvE, 77,
  036307

\bibitem[1980]{KR80} Krause, F. \& R\"adler, K.-H. 1980,
  \emph{Mean-field Magnetohydrodynamics and Dynamo Theory} (Pergamon
  Press, Oxford)

\bibitem[2005]{MGM05} Mininni, P. D., G\'omez, D. O., \& Mahajan, S. M.
  2005, ApJ, 619, 1019

\bibitem[2009]{MKTB08} Mitra, D., K\"apyl\"a, P. J., Tavakol, R. \&
  Brandenburg, A. 2009, \aap, 495, 1

\bibitem[1978]{M78} Moffatt, H. K. 1978, \emph{Magnetic field
    generation in electrically conducting fluids} (Cambridge
  Univ. Press, Cambridge)

\bibitem[1992]{NBJRRST92} Nordlund, \AA, Brandenburg, A., Jennings,
  R. L., et al. 1992, \apj, 392, 647

\bibitem[2003]{O03} Ossendrijver, M. 2003, A\&A Rv., 11, 287

\bibitem[2001]{OSB01} Ossendrijver, M., Stix, M. \& Brandenburg,
  A. 2001, \aap, 376, 726

\bibitem[2002]{OSRB02} Ossendrijver, M., Stix, M., R\"udiger, G. \&
  Brandenburg, A. 2002, \aap, 394, 735

\bibitem[1979]{P79} Parker, E. N. 1979, \emph{Cosmical Magnetic
    Fields: Their Origin and Their Activity} (Clarendon Press, Oxford
  \& NY)

\bibitem[2008]{P08} Pipin, V. V. 2008, GAFD, 102, 21

\bibitem[2009]{PS08} Pipin, V. V. \& Seehafer, N. 2009, \aap, 493, 819

\bibitem[2007]{P07} Proctor, M. R. E. 2007, MNRAS, 382, 39

\bibitem[1968]{R68} R\"adler, K.-H. 1968, Z. Naturforsch., 23a, 1851

\bibitem[1969]{R69} R\"adler, K.-H. 1969,
  Monatsber. Dtsch. Akad. Wiss. Berlin, 11, 194

\bibitem[1980]{R80} R\"adler, K.-H. 1980, AN, 301, 101

\bibitem[2006]{RS06} R\"adler, K.-H. \& Stepanov, R. 2006, PhRvE, 73,
  056311

\bibitem[2003]{RKR03} R\"adler, K.-H., Kleeorin, N. \& Rogachevskii,
  I. 2003, GAFD, 97, 249

\bibitem[2003]{RK03} Rogachevskii, I. \& Kleeorin, N. 2003,
  PhRvE, 68, 036301

\bibitem[2004]{RK04} Rogachevskii, I. \& Kleeorin, N. 2004,
  PhRvE, 70, 046310

\bibitem[2002]{RJ02} Rotvig, J. \& Jones, C. A. 2002, PhRvE, 66, 056308

\bibitem[2004]{RH04} R\"udiger, G. \& Hollerbach, R. 2004,
  \emph{The magnetic Universe}, Wiley-VCH, Weinheim

\bibitem[2006]{RK06} R\"udiger, G. \& Kitchatinov, L.L. 2006,
  AN, 327, 298

\bibitem[2005]{SRSRC05} Schrinner, M., R\"adler, K.-H., Schmitt, D.,
  et al. 2005, AN, 326, 245

\bibitem[2007]{Shea07} Schrinner, M., R\"adler, K.-H., Schmitt, D.,
  et al. 2007, GAFD, 101, 81

\bibitem[2008]{SBS08} Sur, S., Brandenburg, A. \& Subramanian, K.
  2008, MNRAS, 385, L15

\bibitem[1998]{TBCT98} Tobias S. M., Brummell, N. H., Clune, Th. L. \&
  Toomre, J. 1998, \apj, 502, L177

\bibitem[2001]{TBCT01} Tobias S. M., Brummell, N. H., Clune, Th. L. \&
  Toomre, J. 2001, \apj, 549, 1183

\bibitem[2008]{TCB08} Tobias, S. M., Cattaneo, F. \& Brummell,
  N. H. 2008, \apj, 685, 596

\bibitem[1992]{VC92} Vainshtein, S. I. \& Cattaneo, F.  1992, \apj,
  393, 165

\bibitem[1997]{VB97} Vishniac, E. T. \& Brandenburg, A. 1997, \apj,
  475, 263

\bibitem[2001]{VC01} Vishniac, E. T. \& Cho, J. 2001, \apj,
  550, 752

\bibitem[2008a]{YHSea08a} Yousef, T. A., Heinemann, T.,
  Schekochihin, A. A., et al. 2008a, PhRvL, 100, 184501

\bibitem[2008b]{YHSea08b} Yousef, T. A., Heinemann, T.,
  Rincon, F., et al. 2008b, AN, 329, 737

\bibitem[2003]{ZR03} Ziegler, U. \& R\"udiger, G. 2003, \aap, 401, 433

\end{thebibliography}
\end{document}